\documentclass[12pt,preprint,epsfig,color]{aastex}
\begin{document}

\title{Cloning Dropouts : Implications for Galaxy Evolution at High Redshift}

\author{Rychard Bouwens\altaffilmark{1,3}}
\author{Tom Broadhurst\altaffilmark{2,4}}
\author{Garth Illingworth\altaffilmark{1,5}}
\altaffiltext{1}{Astronomy Department,
    University of California,
    Santa Cruz, CA 95064}
\altaffiltext{2}{Racah Institute of Physics, Hebrew University,
       Jerusalem, Israel}
\altaffiltext{3}{bouwens@ucolick.org}
\altaffiltext{4}{tjb@ilios.fiz.huji.ac.il}
\altaffiltext{5}{gdi@ucolick.org}

\begin{abstract}

The evolution of high redshift galaxies in the two Hubble Deep Fields,
HDF-N and HDF-S, is investigated using a cloning technique that
replicates $z\sim 2-3$ $U$ dropouts to higher redshifts, allowing a
comparison with the observed $B$ and $V$ dropouts at higher redshifts
($z\sim 4-5$).  We treat each galaxy selected for replication as a set
of pixels that are $k$-corrected to higher redshift, accounting for
resampling, shot-noise, surface-brightness dimming, and the
cosmological model.  We find evidence for size evolution (a
1.7$\times$ increase) from $z\sim5$ to $z\sim2.7$ for flat geometries
($\Omega_{M}+\Omega_{\Lambda}=1.0$).  Simple scaling laws for this
cosmology predict that size evolution goes as $(1+z)^{-1}$, consistent
with our result.  The UV luminosity density shows a similar increase
(1.85$\times$) from $z\sim5$ to $z\sim2.7$, with minimal evolution in
the distribution of intrinsic colors for the dropout population.  In
general, these results indicate less evolution than was previously
reported, and therefore a higher luminosity density at $z\sim4-5$
($\sim50$\% higher) than other estimates.  We argue the present
technique is the preferred way to understand evolution across samples
with differing selection functions, the most relevant differences here
being the color cuts and surface brightness thresholds (e.g., due to
the $(1+z)^4$ cosmic surface brightness dimming effect).

\end{abstract}


\section{Introduction}

The results from several UV-optical selected samples of galaxies have
recently been pieced together to construct a history of star formation
over a wide range in redshift (Madau et al.\ 1996; Madau, Pozzetti, \&
Dickinson 1998).  A sharp rise in the average star-formation rate is
inferred over the interval 0$<$z$<$1, owing to an increasing incidence
of starburst activity (Broadhurst, Ellis, \& Shanks 1988, Lilly et
al.\ 1996; Glazebrook et al.\ 1994; Cowie et al.\ 1999).  At much
higher redshift, where a well defined sample of field galaxies can be
constructed (Steidel et al.\ 1999) the UV-luminosity density saturates
somewhere before $z\sim3$, placing the peak star-formation rate at a
modest redshift of $z\sim2$ (Madau et al.\ 1996; Madau et al.\ 1998;
Ben{\'i}tez et al.\ 1998).

A comparison of the $U$ and $B$-band dropout galaxies in the HDF North
initially led to claims for a marked decline in the integrated
star-formation rate at $z>2.5$ (Madau et al.\ 1996), but the
spectroscopic work by Steidel et al.\ (1999) on wide-area
ground-selected $U$ and $B$ dropout samples has demonstrated there to
be only a modest evolution in the integrated UV-density from $z\sim3$
to $z\sim4$.  While Steidel et al.\ (1999) speculated that the
relatively small number of $B$ dropouts in the HDF North was just a
downward statistical fluctuation not atypical for such a narrow field,
analyses of the HDF South (Casertano et al.\ 2000) revealed a
similarly large decline, suggesting the need to make a more careful
comparison of these two results.

Detailed observations of the early evolution of galaxies at high
redshift is very important for developing a more concrete
understanding of galaxy formation and for examining the way structure
forms in general.  In this paper, we take a careful model-independent
look at the differential evolution across the high redshift $U$, $B$,
and $V$ dropout populations in the HDF North and South, to thoroughly
address the evolution of the statistical properties of high-z galaxies
over a wide range of redshift, i.e., 2$<$z$<$6.  We replicate the $U$
dropout galaxies to higher redshift, $k$-correcting individual pixels
and using the product of the cosmological volume and a variant of the
space density $1/V_{max}$ to define the number of galaxies.  Care is
taken to account for the instrumental and cosmological transformations
required to project objects to higher redshift, so that the result is
a fully realistic 'no-evolution' simulation from which $B$ and $V$
dropouts can be selected.  Note that our procedure is an improvement
over that used in our earlier work on the general evolutionary
properties of faint field galaxies (Bouwens, Broadhurst, \& Silk
1998a, hereafter denoted BBSI).

We begin by discussing the definition of our high redshift $U$, $B$
and $V$ dropout samples ($\S2$).  In $\S3$ we present our basic
results, in $\S4$ we illustrate how these results might depend upon
geometry or spectral template set, and in $\S5$, we discuss these
results in the larger scope of galaxy formation and
evolution.  Finally, we summarize our findings in $\S6$.   Note that
we frequently denote the HDF $F300W$, $F450W$, $F606W$, $F814W$,
$F110W$, and $F160W$ bands as $U_{300}$, $B_{450}$, $V_{606}$,
$I_{814}$, $J_{110}$, and $H_{160}$, respectively, we assume
$\Omega_M=0.3$, $\Omega_{\Lambda}=0.7$, and we adopt $H_0 = 70\,
\textrm{km/s/Mpc}$ to simplify the expression of scaled
quantities.\footnote{Note that we explore possible sensitivities to
cosmology in \S4.}\footnote{Note that in our analysis the chosen
Hubble constant only has an effect on the units in which the derived
LFs and cosmic star formation rates are expressed.}

\section{High-redshift Selection Criteria}

We make use of the HDF North and South WFPC2 UBVI images (Williams et
al.\ 1996; Casertano et al.\ 2000) and the raw NICMOS JH images of the
HDF-North reduced by Dickinson et al.\ (1999).  After reducing the
NICMOS images, we registered them to coincide with the optical WFPC2
images using our own registration code.  We only consider the central
clean, relatively uniform regions of the WF CCDs, both in the North
and South, each covering roughly 15500 arcsec$^2$.  We degrade the
images of the HDF North slightly to match the depth of the HDF
South--which is slightly shallower by $\sim0.1-0.2$ mags depending on
the passband.  We exclude the PC and noisier edge regions because of
the difficulties in dealing with such heterogenous selection criteria.

Perhaps the most obvious way of selecting high-redshift samples is the
direct approach: to consider only those galaxies with spectroscopic
and photometric redshifts lying within a specific range.
Unfortunately, the blind application of such an approach--in
particular, with regard to photometric redshifts--results in a sample
with a fair number of low redshift contaminants.  Better perhaps to be
a little more conservative and only select objects with colours known
almost certainly to lie in a specific redshift range, well tested by
spectroscopy.  Such selection is possible because high redshift blue
continuum-dominated galaxies occupy a particularly unique region in
color-color space because of the strong Lyman-continuum break at 912
$\textrm{\AA}$ (Meier 1976; Cowie \& Lilly 1988; Guhathkurta, Tyson,
\& Majewski 1990; Steidel \& Hamilton 1992; Steidel \& Hamilton 1993)
and because of an increasingly strong break at higher redshift caused
by the intervening Lyman-alpha forest eating into the spectrum
shortward of 1216 $\textrm{\AA}$ (Madau 1995).

To help define the regions in colour-colour space where high-redshift
objects lie, we model Lyman-dropout objects as young starbursts whose
spectral variations can largely be explained by a range of dust
content.  This choice is motivated by the apparent similarities in
surface brightness and appearance of high redshift galaxies to local
starburst galaxies (Meurer et al.\ 1996; Hibbard \& Vacca 1997) and by
fits performed by Sawicki \& Yee (1997) and Papovich, Dickinson, \&
Ferguson (2001) indicating that the stellar populations of the
Lyman-break objects are very young.  Furthermore, it has been shown
(Calzetti et al.\ 1994) that much of the scatter in the spectra of
starbursts can be accounted for by varying the overall extinction.
Accordingly, we use a single starburst spectra to generate a set of
spectral templates by applying a range of extinctions to some base
SED, which we take to be the solar metallicity 1 Gyr continuous star
formation model used by Steidel et al.\ (1999) to facilitate
comparisons with that work.  We use the Bruzual \& Charlot (2000)
spectrophotometric tables to calculate this base spectrum.  Hereafter,
we abbreviate the associated template set as BC.  We include the
Lyman-alpha continuum and forest absorption according to the
prescription given in Madau (1995) using high quality QSO spectra in
Haardt \& Madau (1996).

We adopt the well-explored $U$-dropout selection criteria of Madau et
al.\ (1996) to produce a sample of objects in the HDFs with
$z\sim2-3.5$: $(B_{450}-I_{814})_{AB} < 1.5$, $(U_{300}-B_{450})_{AB}
> 1.3$, $(U_{300}-B_{450})_{AB} > 1.2 + (B_{450} - I_{814})_{AB}$,
$B_{450,AB} < 26.8$.  We have added a $B_{450,AB} > 22.5$ selection
criterion to exclude low-redshift ellipticals and have required the
$I_{814}$ stellarity parameter (SExtractor, Bertin \& Arnouts 1996) to
be less than 0.85, this serving to exclude most point-like stars from
our catalogs.

For the $B$-dropout selection criteria ($z\sim3.5-4.5$), our limits
differ somewhat from Madau et al.\ (1996): $(B_{450} - V_{606})_{AB} >
1.4$, $(B_{450} - V_{606})_{AB} > 3.8 (V_{606} - I_{814})_{AB} -
1.07$, $V_{606,AB} < 27.7$, $V_{606,AB} > 22.5$, and a stellar
parameter less than 0.85.  We restricted our sample to $V_{606}$
magnitudes brighter than 27.7 to avoid selecting objects that are
intrinsically fainter than are selectable in our $U$-dropout sample.

To select $V$-band dropouts ($z\sim4.5-5.5$), we use the following
selection criterion: $(V_{606} - I_{814})_{AB} > 1.5$, $(V_{606} -
I_{814})_{AB} > 3.8 (I_{814} - H_{160})_{AB} - 1.54$, $I_{814,AB} <
27.6$, $I_{814,AB} > 24$.  We restricted our sample to $I_{814,AB}$
magnitudes brighter than 27.6 to avoid selecting objects that are
intrinsically fainter than are selectable in our $U$-dropout
sample.  For the $U$, $B$, and $V$ dropouts, Figures 1-3 illustrate
the tracks that different starburst templates (e.g., $E(B-V) = 0.0,
0.2, 0.4$) make in their respective colour-colour diagrams.

\begin{figure}
\epsscale{0.95}
\plotone{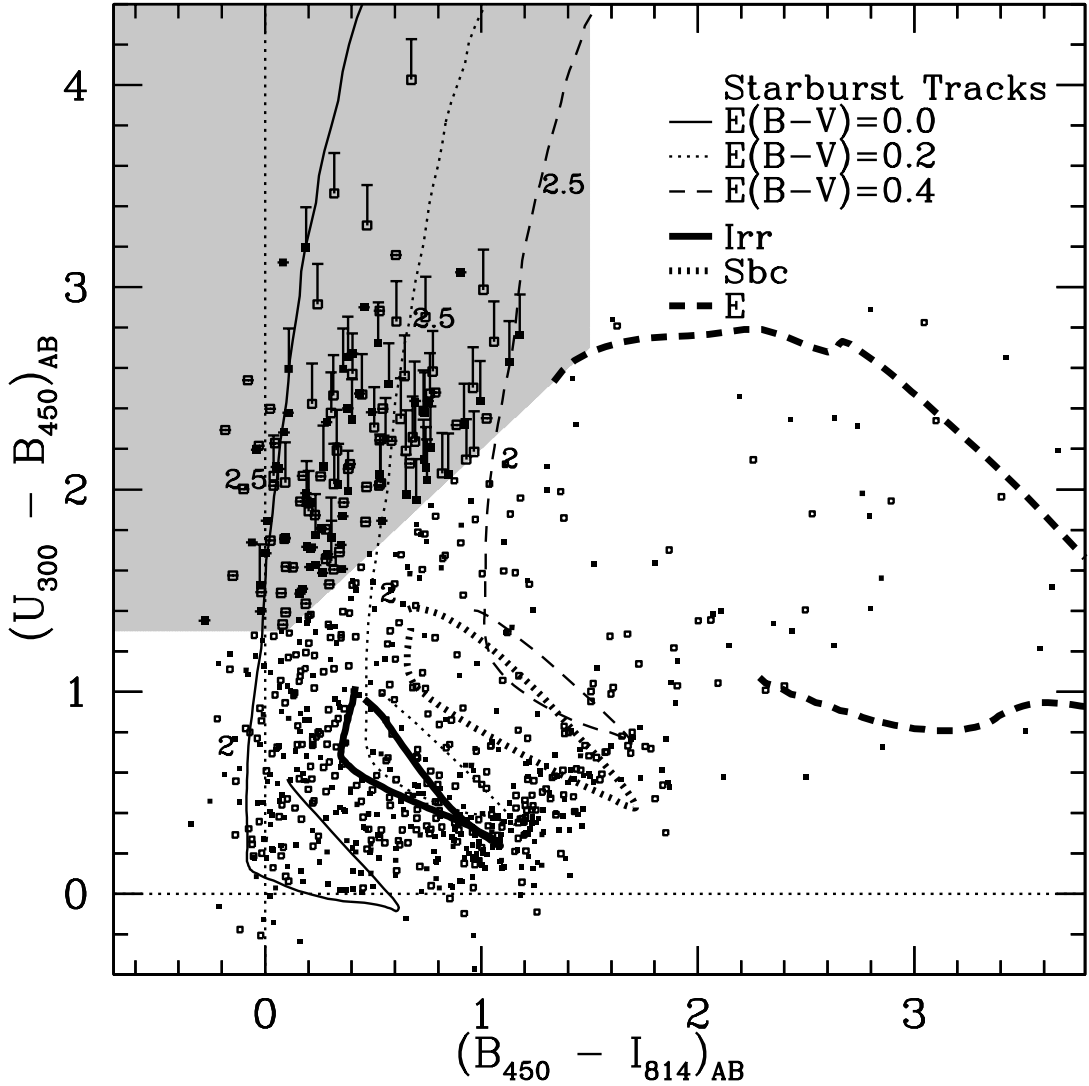}
\caption{$(U_{300}-B_{450})_{AB}$/$(B_{450} - I_{814})_{AB}$
colour-colour diagram illustrating the position of our $U$-dropout
sample (shaded region) relative to the photometric sample as a
whole.  Tracks for a $10^9$ year starburst with various amounts of
extinctions have been included to illustrate both the typical
redshifts and SED types included in the selection window.  The
low-redshift ($0<z<1.2$) tracks for typical E, Sbc, and Irr spectra
have been included as well to illustrate the region in colour-colour
space where possible contaminants might lie.  Solid (open) squares
indicate objects from the HDF North (HDF South).  Larger squares
indicate objects found in our sample.  Error bars represent 1.5
$\sigma$ limits.}
\end{figure}

\begin{figure}
\epsscale{0.95}
\plotone{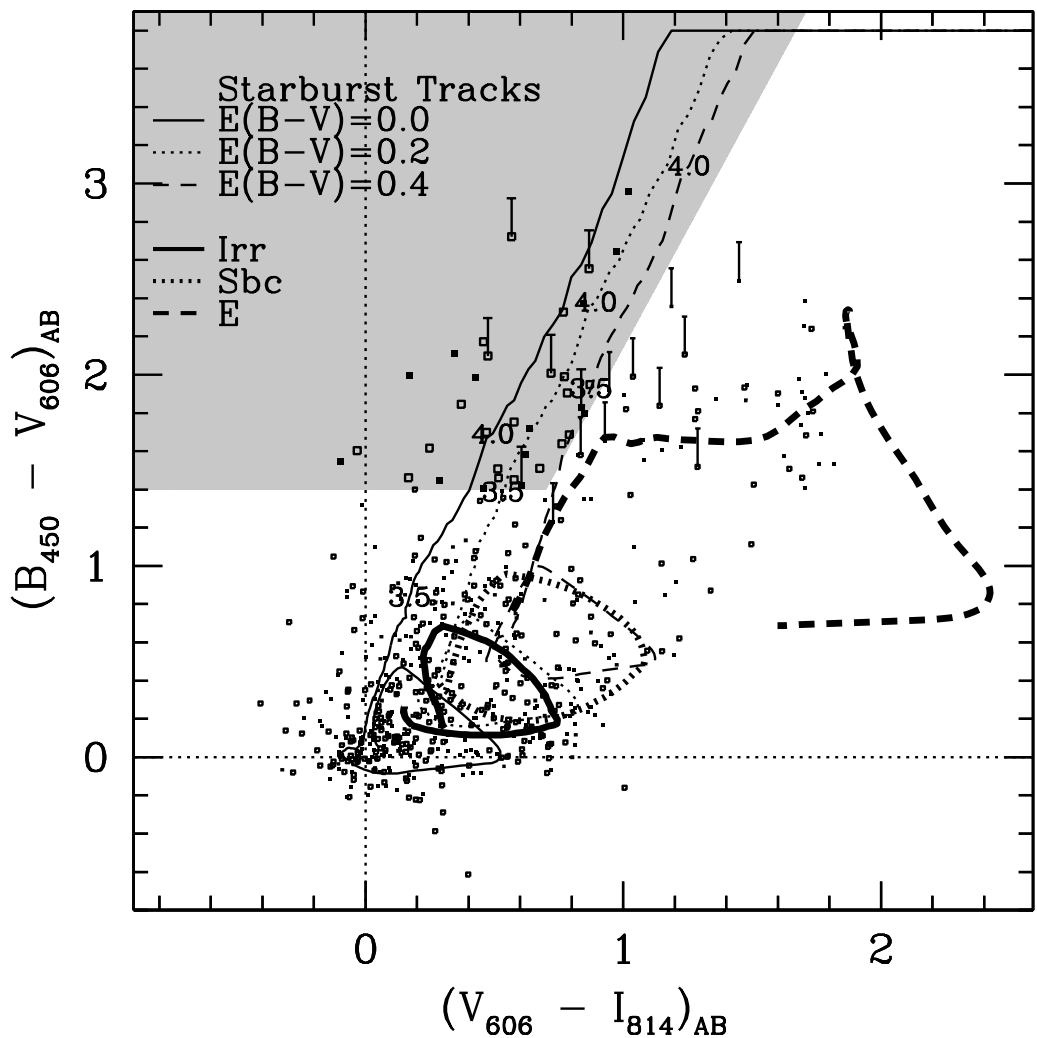}
\caption{$(B_{450}-V_{606})_{AB}$/$(V_{606} - I_{814})_{AB}$
colour-colour diagram illustrating the position of our $B$-dropout
sample (shaded region) relative to the photometric sample as a
whole.  Otherwise the same as Figure 1.}
\end{figure}

\begin{figure}
\epsscale{0.95}
\plotone{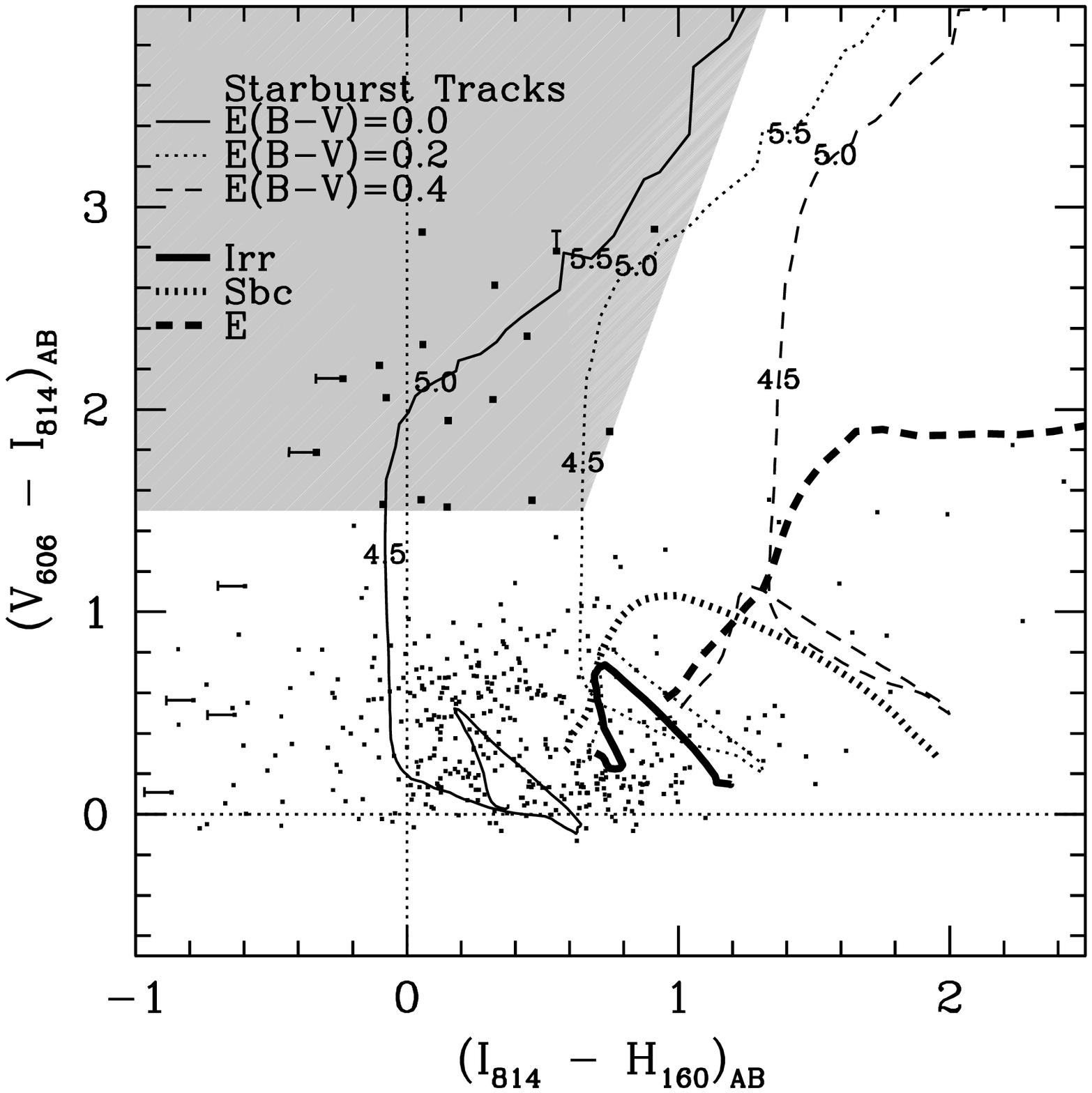}
\caption{$(V_{606}-I_{814})_{AB}$/$(I_{814} - H_{160})_{AB}$
colour-colour diagram illustrating the position of our $V$-dropout
sample (shaded region) relative to the photometric sample as a
whole.  Otherwise the same as Figure 1.}
\end{figure}

Due to the small number of objects ($\sim19$) in the above $V$-dropout
sample, we consider an alternative $V$-dropout sample, with very red
$(V_{606}-I_{814})_{AB}>1.8$ colors and no requirement on the
optical-to-near-infrared color.  This enables us to include the HDF
South data where no deep space-based near-infrared images exist.
While one might worry about the presence of lower redshift
contaminants like EROs (Extremely Red Objects) in such a sample, the
spectral slope redward of the break (e.g., the
$(I_{814}-H_{160})_{AB}$ color) for similarly red objects in the HDF
North tends to be relatively flat, indicating that most faint red
objects are indeed at high-z ($z\sim4-6$), and so the contamination is
not very large.  We call this sample the optical $V$-dropout sample
(``V-dropout (Opt)'') to distinguish it from the $V$-dropout sample
described in the previous paragraph where infrared fluxes are used
(``V-dropout (IR)'').

We have made sure to set up the selection criteria so that the
intrinsic set of objects selected by either our $B$ dropout criterion
or our $V$ dropout criterion--as parametrized by absolute magnitude or
spectral index--are subsets of those selected by our $U$ dropout
criterion.  This is important whenever one projects one sample onto
another for the sake of intercomparison: one needs to insure that the
set of galaxies into which one is mapping, i.e., the range, is
strictly a subset of the galaxies one is mapping, i.e., the domain.
Otherwise, one can not preclude there being some population of
galaxies in the range (e.g., the mapped-into sample) which do not have
duplicates in the domain (the mapped sample).  In the present case,
this would mean making the mistake of comparing a $U$-dropout sample,
defined to include only the bluest U-dropouts ($E(B-V)<0.1$), with $B$
and $V$-dropout samples, defined to include all ranges of intrinsic
$E(B-V)$ reddenings, by projecting the former onto the latter.  The
selection criteria given above were chosen to avoid these type of
difficulties.

\section{Results}

\subsection{Derived Samples}

Using the photometry and sample selection procedure described in
Appendix A, we found 61 and 72 $U$-dropouts for the HDF North and
South, respectively.  We found 15 and 21 $B$-dropouts for the HDF
North and South, respectively.  Finally, we found 19 $V$-dropouts for
the HDF North.  We also found 12 and 6 objects in the HDF North and
South, respectively, with very red $(V_{606}-I_{814})_{[AB}>1.8)$
colors.  We determined the object redshifts photometrically, or
spectroscopically if available, and determined the SED templates that
best-fit their pixel-by-pixel fluxes.  Spectroscopic redshifts are
available for 17 objects from our $U$-dropout sample and are in
excellent agreement with our photometric redshift estimates (see
Figure 4), the overall RMS scatter $\sqrt{\left\langle\left[(\Delta
z)/(1+z)\right]^2\right\rangle}$ being only 0.05.  This being said,
the photometric redshifts do seem to have a small upward bias in
redshift compared to the spectroscopic measures, e.g., $\left\langle
(\Delta z)/(1+z) \right\rangle = 0.02$.  This small bias does not
appear to be a big problem because we were able to independently
replicate all our results (within $\sim10-15$\%) using the Steidel et
al.\ (1999) $z\sim3$ luminosity function.  We remark on this briefly
at the end of \S4.  Figure 5 contrasts the intrinsic SED distribution
we derive with that of Steidel et al.\ (1999).  As detailed in
Appendix A, the intrinsic SED be parametrized in terms of $E(B-V)$ by
applying various amounts of extinction to some base spectrum.
Clearly, our intrinsic SEDs are slightly bluer than those of Steidel
et al.\ (1998).  While this might well indicate slight differences in
the exact shape of the SED templates and possible redshift biases,
they don't result in any large systematics.  We comment on this later
in \S4.  Finally, we provide basic plots of the number counts and
angular size (half-light radii) distributions we obtained for these
samples in Figure 6-10.\footnote{The sizes described here are
half-light radii and are derived by calculating the growth curve as a
function of radius and selecting that radius which contains half the
total light contained within three Kron (1980) radii.}

\begin{figure}
\epsscale{0.95}
\plotone{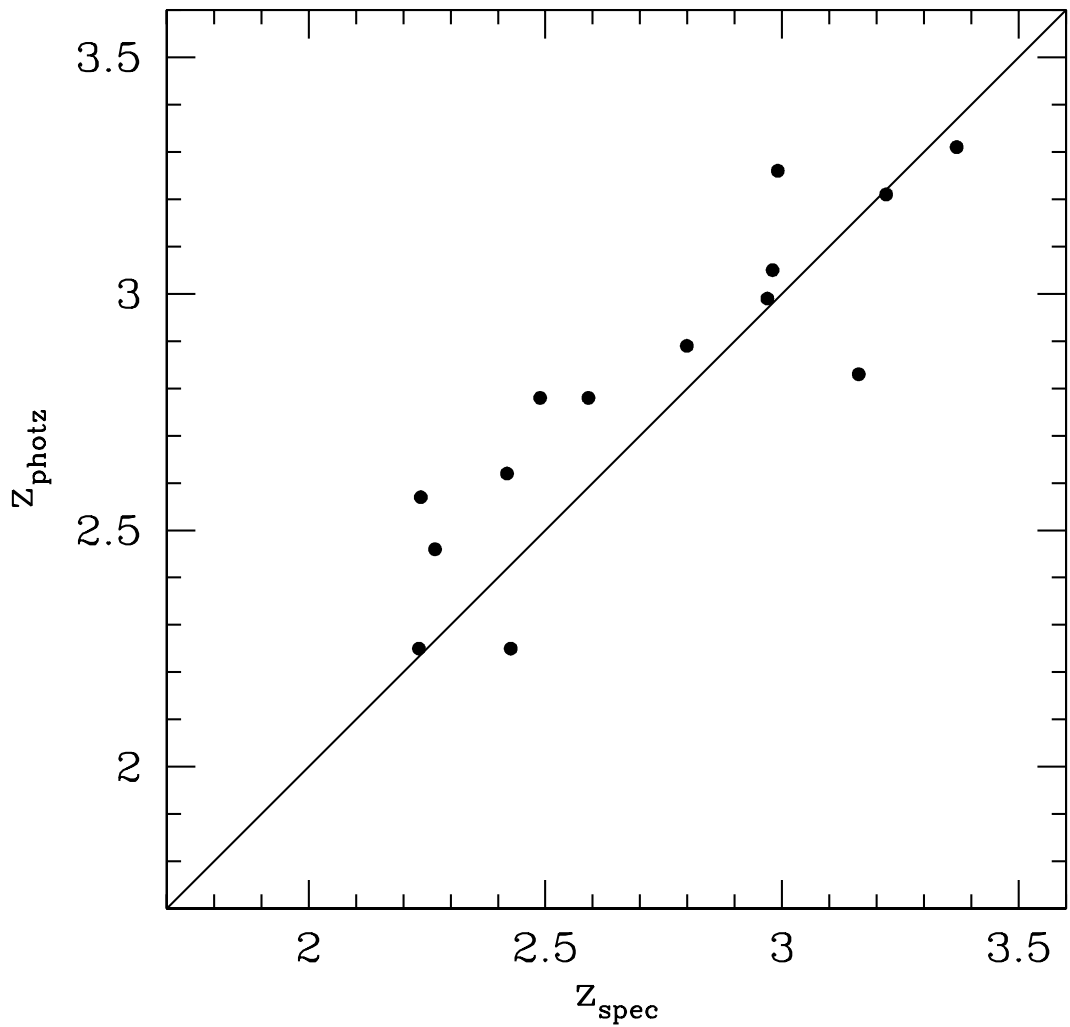}
\caption{The upper panel compares the photometric redshifts we
estimate for the $U$-dropouts with the spectroscopic values.  The
scatter is small, $\sqrt{\left\langle\left[(\Delta
z)/(1+z)\right]^2\right\rangle} = 0.05$.  There is a small upward bias
in redshift compared to the spectroscopic measures, e.g.,
$\left\langle (\Delta z)/(1+z) \right\rangle = 0.02$, but this is not
an issue because we were able to independently replicate all our
results (within $\sim10-15$\%) using the Steidel et al.\ (1999)
$z\sim3$ luminosity function (\S4).}
\end{figure}

\begin{figure}
\epsscale{0.95}
\plotone{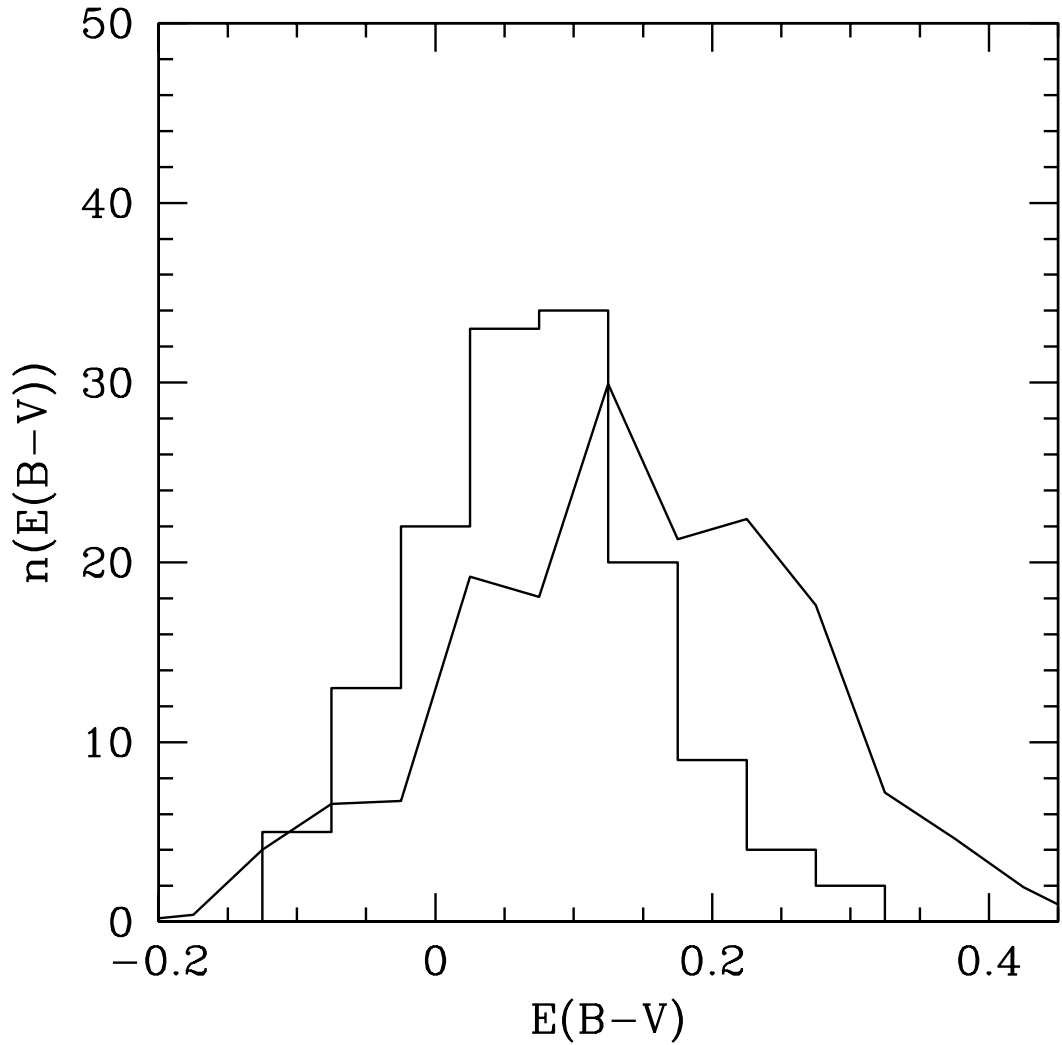}
\caption{Comparison of the $E(B-V)$ distribution recovered for our
$U$-dropout sample (histogram) with Steidel et al.'s (1999)
determination (line) (see Figure 6 from that paper.)  The relative
normalization is set so as to allow comparison between Steidel's
sample and our HDF objects.  The offset does not appear to be
significant, particularly because we are able to reproduce all the
latter results using both the $E(B-V)$ distribution shown above and
the Steidel et al.\ (1999) luminosity function (\S4).  Negative values
of $E(B-V)$ are used here for similarity with Steidel et al.\ (1999)
since they provide a convenient way of representing templates bluer
than our base spectral template.}
\end{figure}

\begin{figure}
\epsscale{0.95}
\plotone{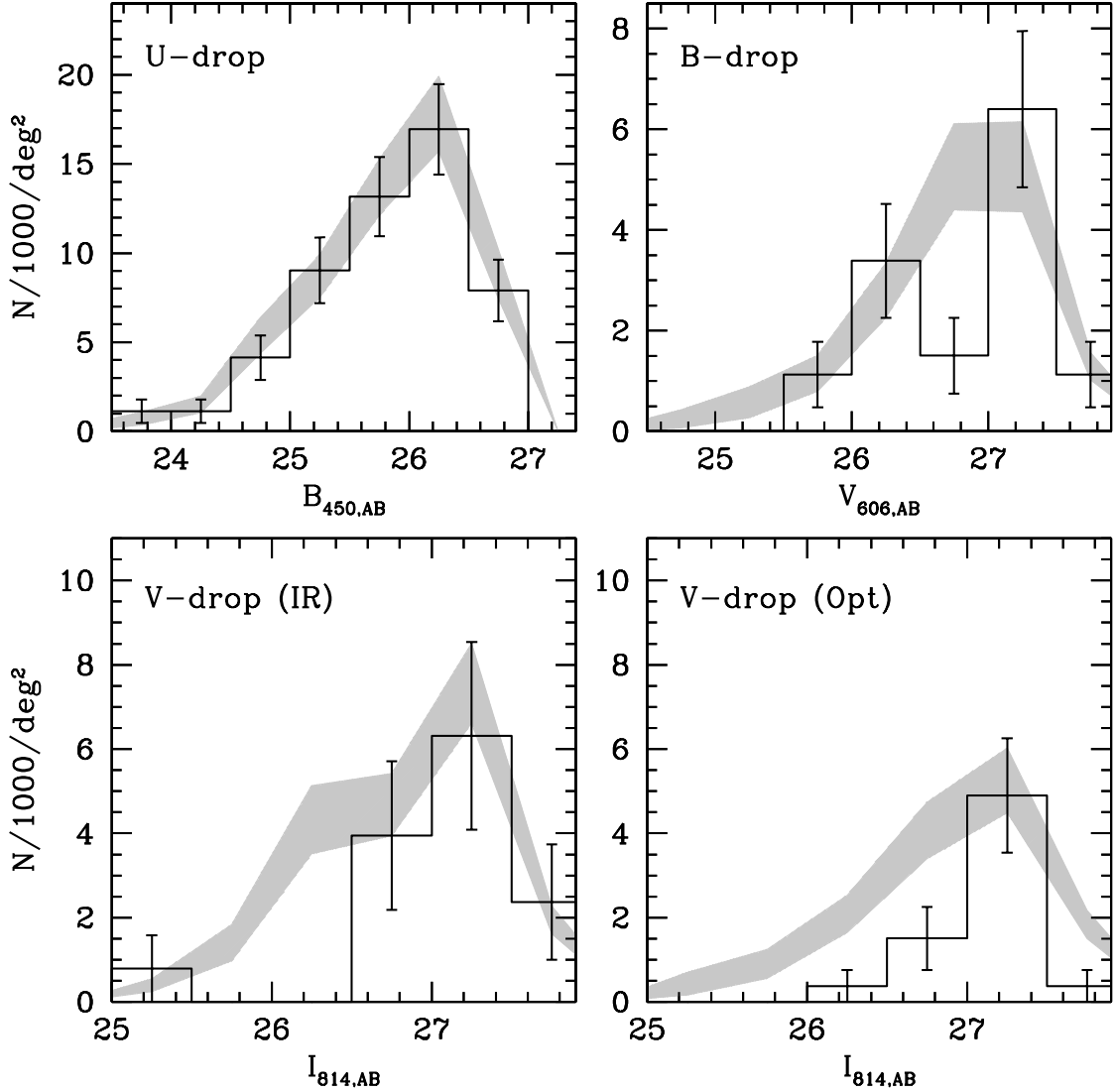}
\caption{Comparison of the number counts for the $U_{300}$, $B_{450}$,
and $V_{606}$ dropouts observed in the HDF (histogram) with the
no-evolution expectations based upon our $U$-dropout sample (shaded
regions).  Definitions of all the dropout samples, including the two
$V$-dropout samples are given in \S2.  Note the good agreement between
the distribution recovered from the $U$-dropout samples (in the
upper-left panel) and the cloning simulations derived from them (see
\S3.4).}
\end{figure}

\begin{figure}
\epsscale{0.95}
\plotone{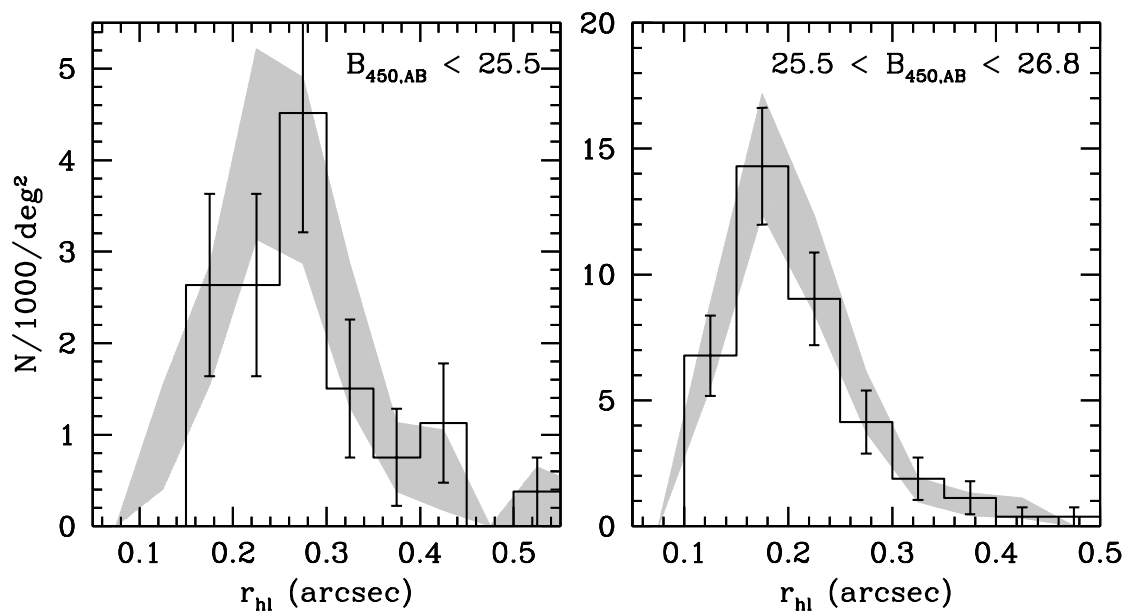}
\caption{Comparison of the half-light radius distributions for the
$U_{300}$ dropout sample with the cloning simulations derived from
them for two different magnitude intervals using the $\Omega_{M}=0.3$,
$\Omega_{\Lambda}=0.7$ geometry.  Excellent agreement between the
simulations and observations points toward a general self-consistency
in our procedure (\S3.4).}
\end{figure}

\begin{figure}
\epsscale{0.95}
\plotone{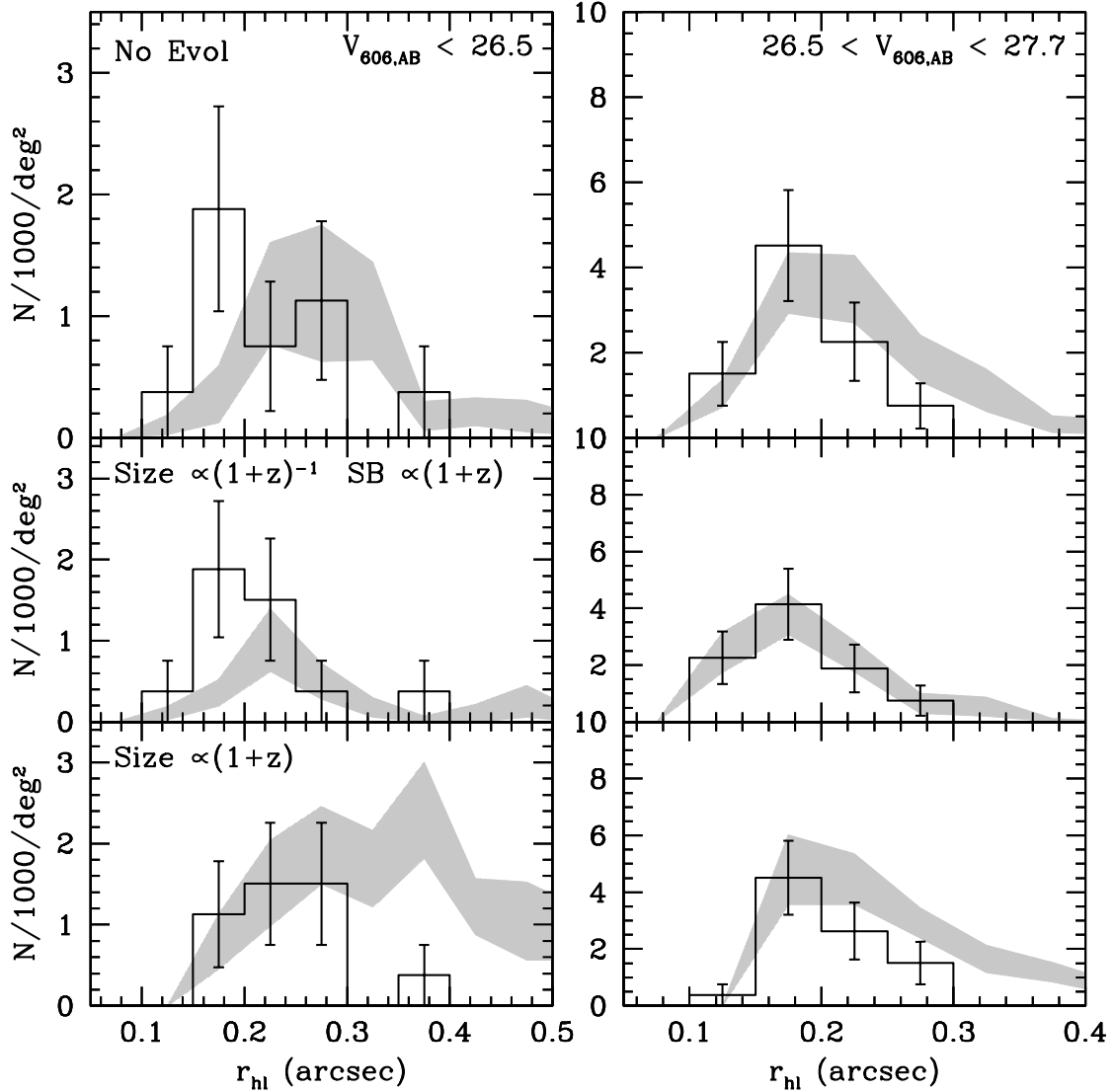}
\caption{Similar to Figure 7, but for the $B_{450}$-dropouts.  The
shaded regions indicate the expected values based upon the $U$-dropout
sample.  The predictions of a no-evolution model are presented in the
top panel, those for our preferred size-evolution model with mild
surface brightness evolution where size scales as $(1+z)^{-1}$ are
presented in the middle panel, and those of a constant surface
brightness size evolution model where sizes scale as $(1+z)$ are given
in the bottom panel.  In both magnitude intervals, the observed
angular size distribution is somewhat smaller than that based upon a
no-evolution projection of the $U$-dropout population.  Note that
because the effective PSF of our $z\sim2-3$ templates is typically
larger than for the observations at $z\sim4$ (see \S3.3 for a more
detailed explanation), it is necessary for us to apply varying amounts
of smoothing to the observations before making a comparison with the
lower redshift population.}
\end{figure}

\begin{figure}
\epsscale{0.95}
\plotone{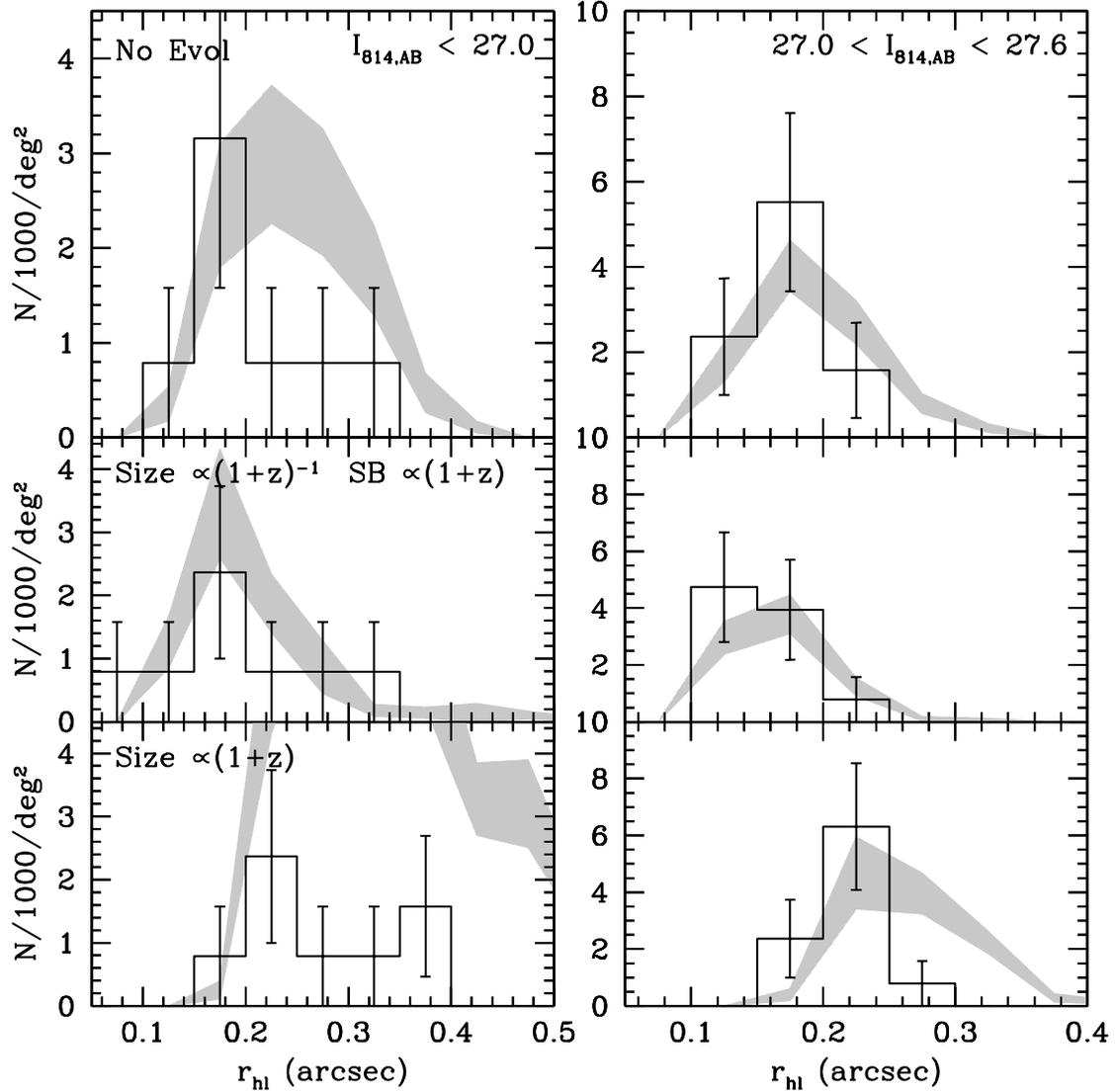}
\caption{Similar to Figure 8, but for $V_{606}$-dropouts selected
using the infrared photometry (``V-drop (IR)'').  The observed angular
size distribution of the bright slice ($I_{814,AB}<27$) (histogram)
are smaller (90\% confidence) than that predicted based on the
$U$-dropout sample ($z\sim2.7$) (top panel), the $(1+z)^{-1}$ model
shown in the middle panel providing the best-fit to the size evolution
observed.  Together with Figure 10, this shows that $UV$ bright
objects are smaller at $z\sim5$ than at $z\sim2.7$.}
\end{figure}

\begin{figure}
\epsscale{0.95}
\plotone{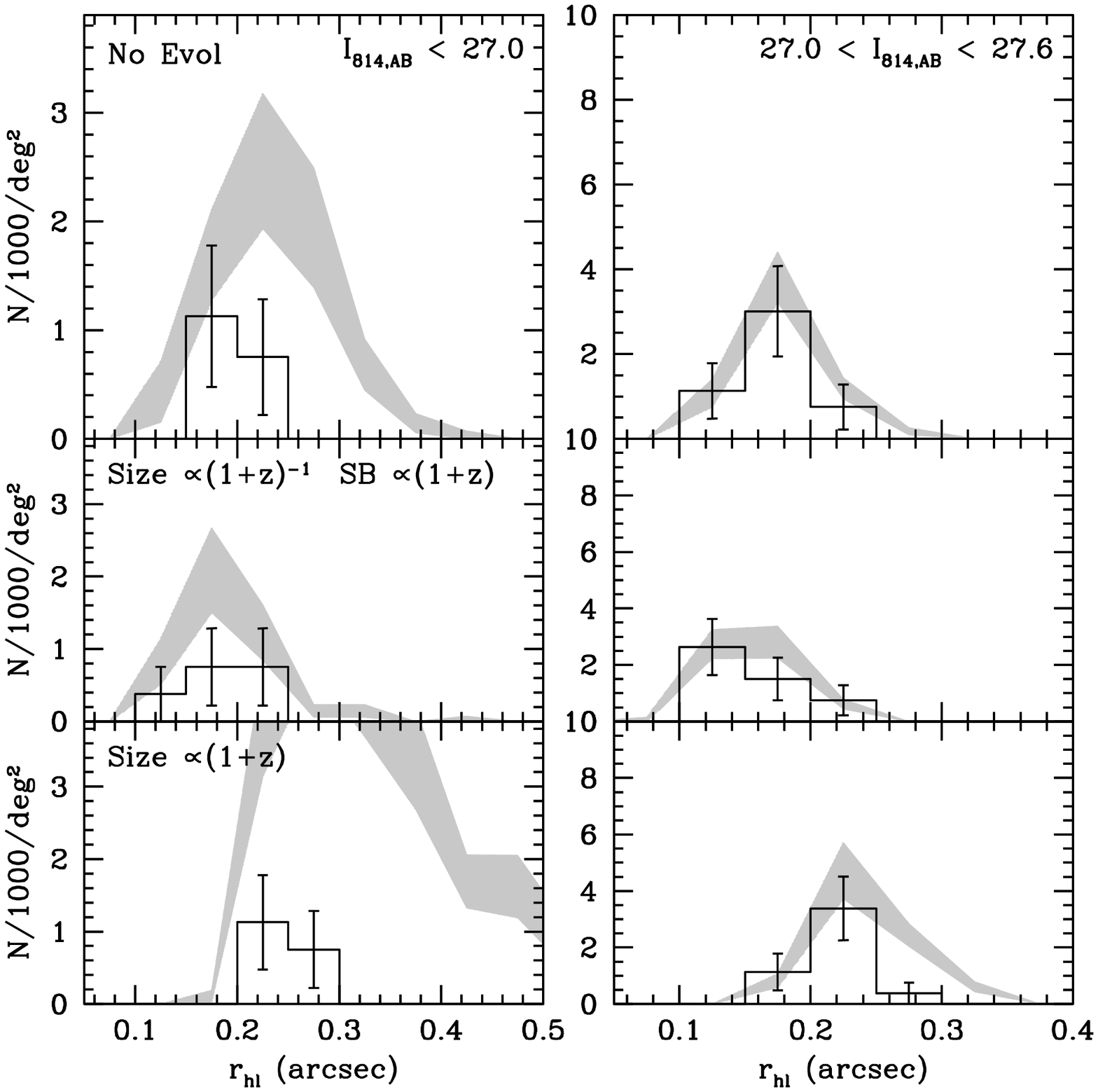}
\caption{Similar to Figure 9, but for the ``optical''
$V_{606}$-dropouts selected from the HDF North and South.  The fact
that the observed angular size distribution (histogram) for the bright
slice ($I_{814,AB}<27$) is shifted toward smaller sizes than that
predicted from the $z\sim2.7$ $U$-dropouts clearly suggests that
galaxies are smaller at $z\sim5$ than at $z\sim2.7$ (86\% confidence).
Together with Figure 9, this shows that galaxies are smaller at
$z\sim5$ than they are at $z\sim2.7$.  In both, the best fit is
provided by the middle panel with a $(1+z)^{-1}$ scaling in size.}
\end{figure}

We determined the volume densities of each object in these base
samples using the procedure given in Appendix C.  Briefly, we
projected the galaxy to all redshifts using the procedure described in
Appendix B, remeasured its properties, and then took its volume
density to be the reciprocal of the effective selection volume.  Using
the volume densities determined in Appendix C, it was straightforward
to derive an estimated luminosity function for both the $U$ and
$B$-dropouts.  Note that we determined $M_{1700,AB}$ magnitudes using
each object's $I_{814}$ magnitude and best-fit pixel-by-pixel SED.  We
include these LFs on Figure 11 in the form of solid and open boxes for
the $U$ and $B$-dropouts, respectively, the error bars representing
one-sigma uncertainties.  We also include various results by Steidel
et al.\ (1999) on that plot, but will also leave a discussion of that
until later (\S5.2).  While the $U$-dropout LF has a normalization
which is roughly $2\times$ higher than the $B$-dropout LF, we will
argue that the intrinsic normalizations are closer than this and the
apparent difference is largely a consequence of the differing
selection effects (\S5.2).

\begin{figure}
\epsscale{0.95}
\plotone{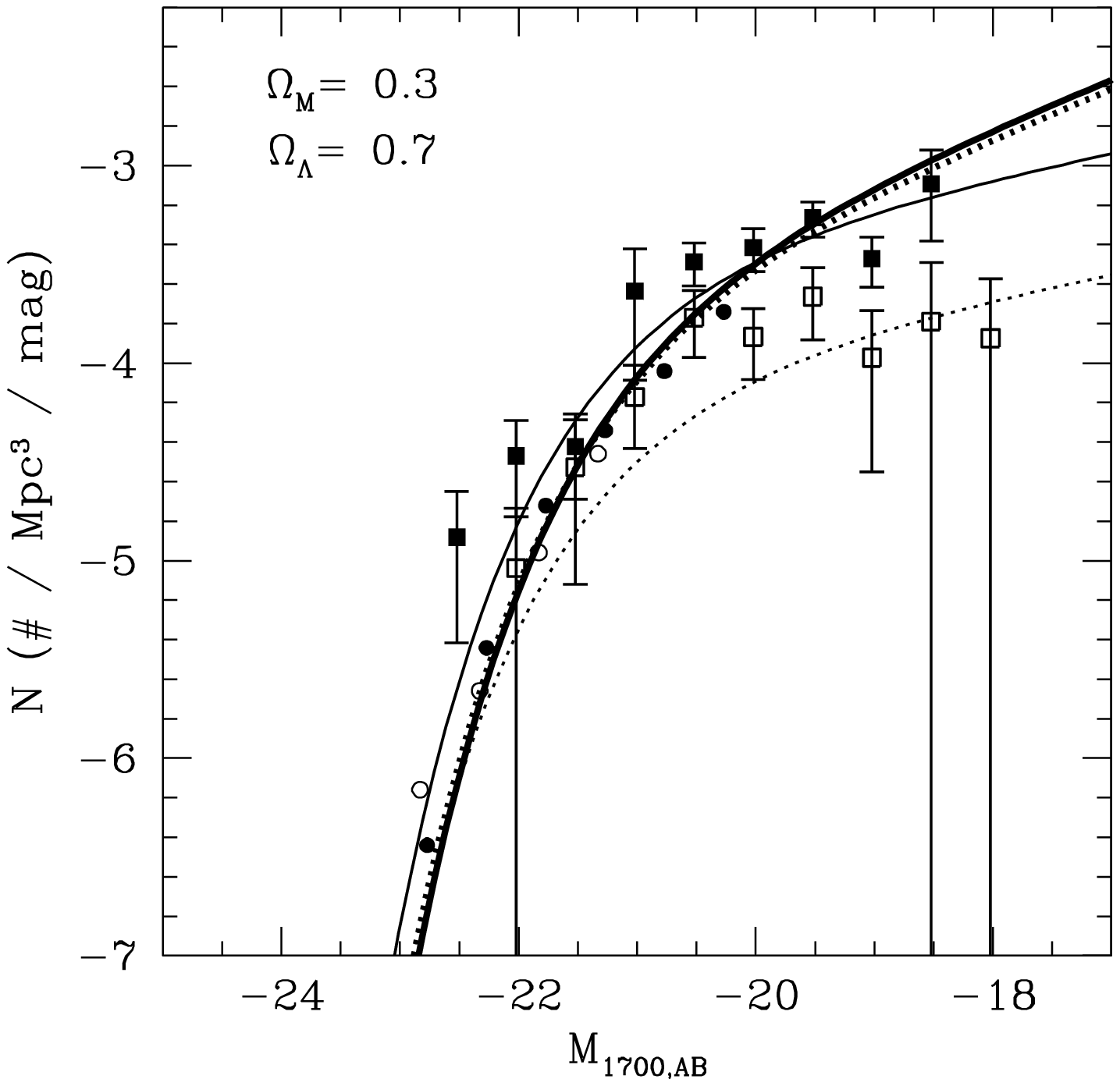}
\caption{Rest-frame ($z=3$) luminosity functions for the U-dropout
sample (filled squares) and the B-dropout sample (open squares) used
in the present study assuming a $\Omega_{M}=0.3$,
$\Omega_{\Lambda}=0.7$ geometry and $H_0 = 70\,
\textrm{km/s/Mpc}$.  Note that the faintest two bins in our LFs suffer
from incompleteness.  Our LF matches Steidel et al.'s (1999) LF at
$z\sim3$ (filled circles; thick solid line), but falls below their
determination at $z\sim4$ (open circles; thick dotted line).  The
Pozzetti et al.\ (1998) LF at $z\sim3$ (thin solid line) and at
$z\sim4$ (thin dotted line) are also shown.}
\end{figure}

\subsection{Sample Fairness}

It is useful to assess the fairness of our samples, especially given
the surprising amount of clustering observed at high redshift (Steidel
et al.\ 1998; Giavalisco et al.\ 1998; Adelberger et al.\ 1998) and
possible systematics in the photometric redshifts we use.  To this
end, we plot $V/V_{max}$ (Schmidt 1968) for our $U$ dropout sample in
Figure 12.  For a homogeneous sample in magnitude and redshift, the
quantity $<V/V_{max}>$ should equal
\begin{displaymath}
0.5 \pm \frac{1}{\sqrt{12N}}
\end{displaymath}
where $N$ is the number of the galaxies in the sample ($N=142$).  Our
samples meet this goal, and even more encouragingly, we find that the
$V/V_{max}$ distributions are relatively flat.

\begin{figure}
\epsscale{0.95}
\plotone{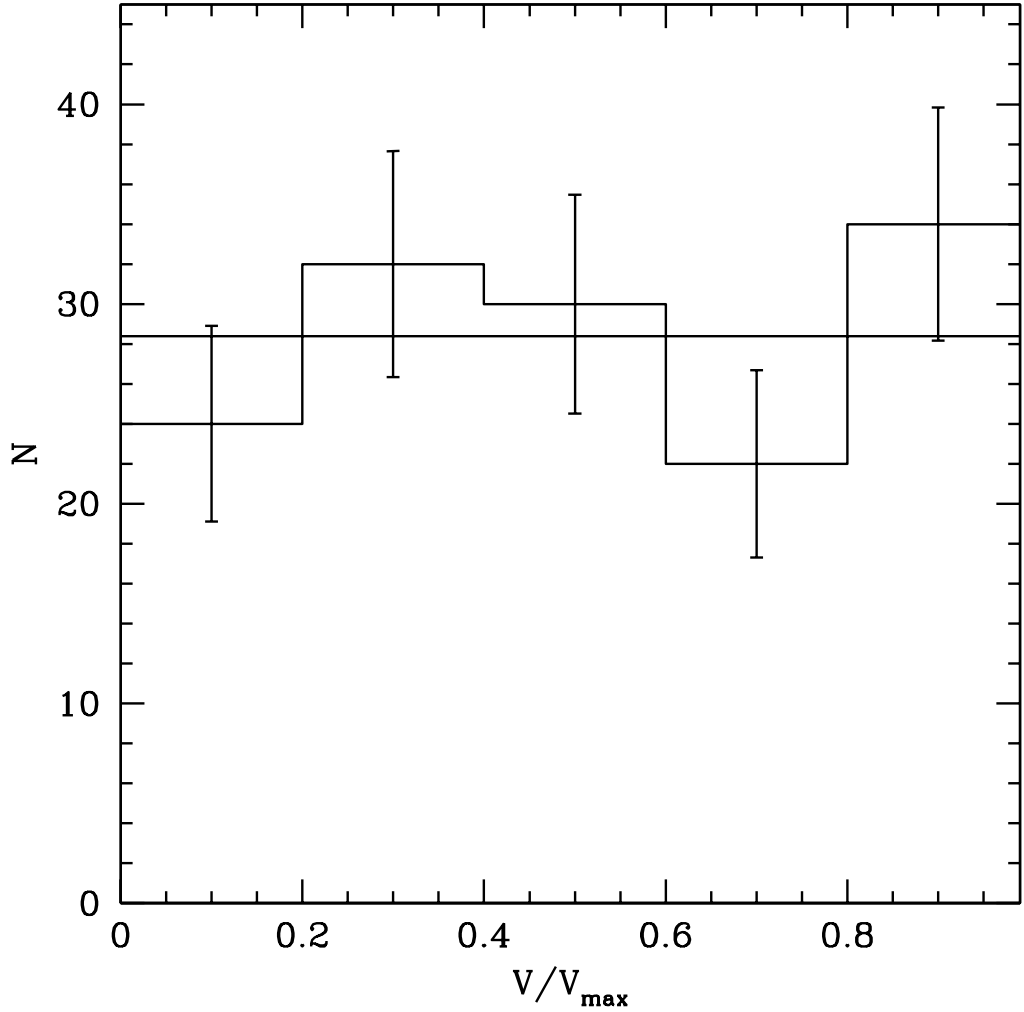}
\caption{$V/V_{max}$ distribution for the $U$ dropout samples for the
$\Omega_{M}=0.3$, $\Omega_{\Lambda}=0.7$ geometry.  The horizontal
line shows the expected value for each bin.  The error bars show the
expected one sigma variations in the expected numbers.  The fact that
the $V/V_{max}$ distribution is flat and here agreement shows that our
$U$ dropout sample is fair.}
\end{figure}

\subsection{Simulating the $U$, $B$, and $V$ Dropout Samples}

It is now completely straightforward to use the number densities
derived to compare the $U$-dropout objects with all the other samples
compiled here, including the $U$ dropout sample itself.  Assuming no
spatial clustering, we use the cosmological volume and object volume
densities to generate Monte-Carlo object catalogs over a given
redshift interval and effective area of $2.5 \times 10^5$
$\textrm{arcsec}^2$.  We simulate the appearance of objects at new
redshifts using the steps outlined in Appendix B.1-B.4, and we measure
their properties using the techniques discussed in Appendix B.5.
Applying the selection criteria relevant to the fiducial sample, we
are able to compile the properties of the projected base sample.  We
compute one-sigma uncertainties on the expected numbers based upon the
number of objects from our input samples which contribute to the
inferred numbers.  Our procedure for deriving error estimates is fully
described in BBSI and is similar to what one would obtain from a
bootstrapping analysis.  In future sections, we often refer to this as
our empirical $U$-dropout model.  It is used almost exclusively to
make predictions.  Before comparing our projected $U$-dropouts with
higher redshift data, we smooth the data slightly so as to match the
projected PSF of $z\sim2-3$ objects at $z\sim4-5$.  This is necessary
because the angular size, and therefore effective PSF, of objects is
smaller at $z\sim2-3$ ($\textrm{FWHM}_{PSF} \sim\,$0.14 arcsec) than
it is when projected to $z\sim3-6$ ($\textrm{FWHM}_{PSF} \sim\,$0.18
arcsec at $z\sim5$) for the $\Omega_{M}=0.3$, $\Omega_{\Lambda}=0.7$
geometry.

\subsection{Simulation Self-Consistency}

Clearly, using the $U$-dropout clones, we should be able to make
Monte-Carlo simulations of the high redshift universe and recover
something similar to the $U$-dropout observations from which the
empirical models were derived.  In Figure 6, we show the observed
number counts for the $U$-dropouts and overplot our cloning
expectations based upon this same $U$-dropout sample.  The shaded
regions show the variation expected due to the finite size of our
input samples.  In Figure 7, we do the same for the angular sizes of
the $U$ dropouts, the histogram indicating the observations and the
shaded regions indicating that expected from cloning the $U$ dropouts.
In all cases, the $U$-dropout samples successfully reproduce the
parent distributions from which they were derived.

\subsection{Number Counts}

Comparing the observed number counts with those predicted from our
empirical $U$-dropout model allows us to test the extent to which the
luminosity, size, and colour distributions of $UV$-bright galaxies
change as a function of redshift.  First, we consider basic number
count predictions.  Figure 6 shows how the observed $B$ and
$V$-dropout number counts compare with those predicted based upon the
$U$ dropout samples (shaded region).  Clearly, we expect more $B$ and
$V$ dropouts based upon the cloned $U$-dropout population than are
actually observed in the HDF fields.  Integrating down the number
counts, one infers there are 40\% less $UV$-bright galaxies at
$z\sim4$ than there are at $z\sim2.7$ and $\sim46$\% less at $z\sim5$
than at $z\sim2.7$.  Obviously, these numbers are slightly different
than those we gave earlier (\S3.1) in comparing the luminosity
functions over this same redshift range, and the reason isn't that
surprising: the $U$-dropout selection criterion includes objects which
aren't included in the $B$-dropout criterion, specifically objects of
lower surface brightness and redder colors.\footnote{Note that this is
preferable to the situation discussed at the end of \S2 where the
higher redshift samples contained objects not contained in the lower
redshift $U$-dropout sample.}  We will discuss these differences a
little more extensively later (\S5.2).

\subsection{Galaxy Sizes}

Next, we proceed to an examination of the angular sizes.  This is
important because it provides crucial new information on the extent to
which galaxies may have evolved in size from $z\sim5$ to $z\sim2.7$,
not discernible using ground-based data.  Figures 7-10 present the
angular size (half-light radii) distributions for the $U$, $B$,
$V$-dropout samples along with a comparison with the predicted
distribution based upon the $U$-dropout population (shaded regions).
Stepping from $z\sim2.7$ to $z\sim4$, a clear size difference is
already apparent, particularly in the brightest magnitude bin
($V_{606,AB} < 26.5$).  This size difference becomes even more obvious
when the comparison is made at $z\sim5$ (both for the optical and
infrared $V$-dropout samples), this difference again being the largest
in the brightest magnitude bin ($I_{814,AB} < 27$).  We illustrate
this difference more graphically by showing a random sample of the
observed $B$-dropouts and those predicted based upon the $U$-dropout
population in Figure 13.   We do a similar thing for the $V$-dropouts
in Figure 14.  Clearly, the observed $V$-dropouts are slightly smaller
on average and more centrally concentrated in surface brightness.

\begin{figure}
\begin{center}
\resizebox{12.2cm}{!}{\includegraphics*[20,21][500,820]{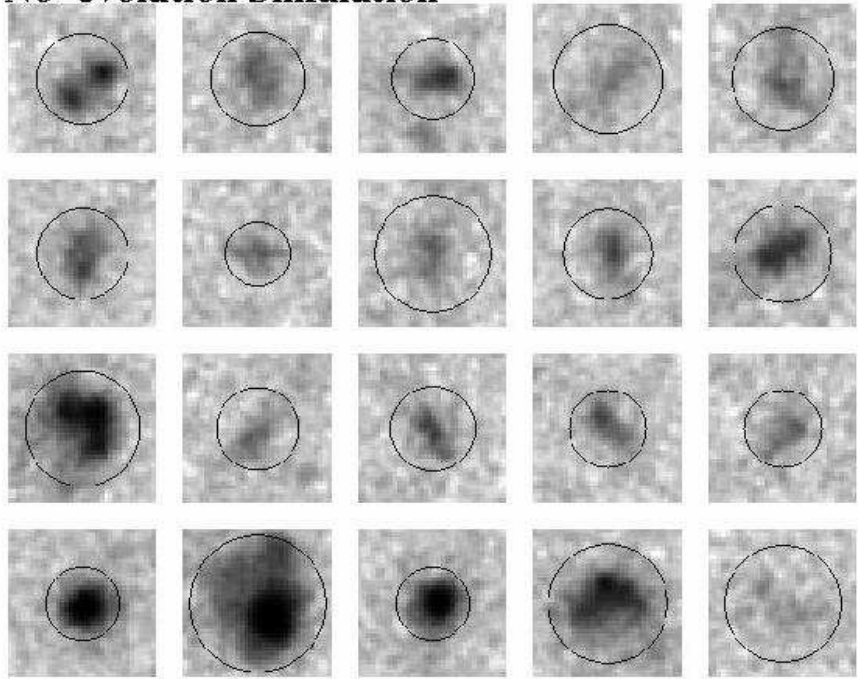}}
\end{center}
\caption{Postage stamp images of the $B$-dropouts ($z\sim4$)
identified in the HDF North and South (upper panel) versus those
expected from a no-evolution extrapolation of our $z\sim2.7$
$U$-dropout sample (lower panel).  Circles demarking twice the
determined half-light radius are included for each object.  As
demonstrated quantitatively in Figure 8, the mean size of the observed
$B$-dropout population is smaller than that predicted based on a
no-evolution extrapolation from the $U$-dropout population and is
quite evident from a comparison of these panels (from the mean size of
the half-light distributions).  Plate scale is 1.36 arcsec.}
\end{figure}

\begin{figure}
\epsscale{0.9}
\begin{center}
\resizebox{12.2cm}{!}{\includegraphics*[20,20][500,820]{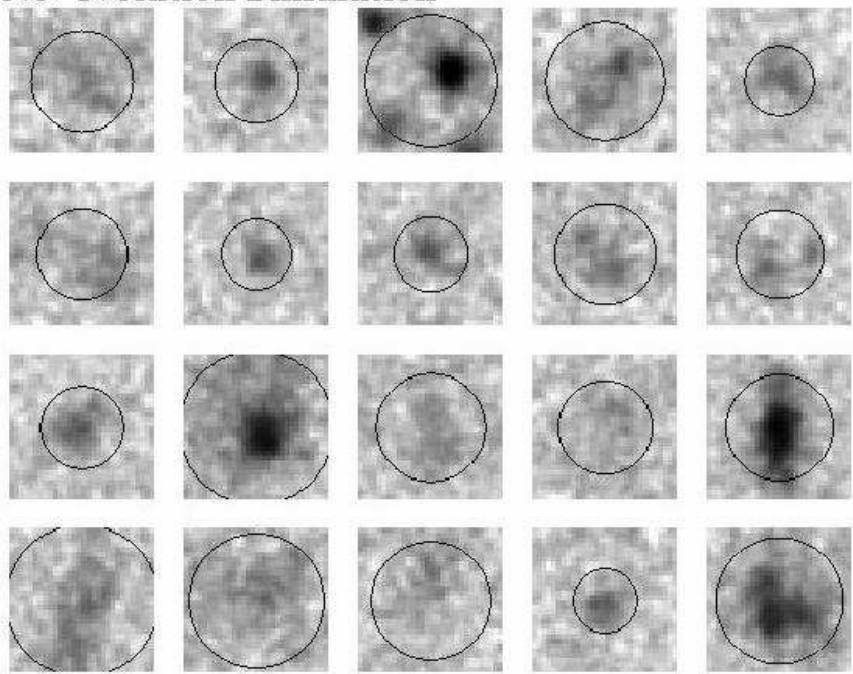}}
\end{center}
\caption{Postage stamp images of the $V$-dropouts ($z\sim5$)
identified in the HDF North and South (upper panel) versus those
expected from a no-evolution extrapolation of our $z\sim2.7$
$U$-dropout sample (lower panel).  All 18 ``optical'' $V_{606}$
dropouts are shown.  Circles demarking twice the determined half-light
radius are included for each object, as in Figure 13.  As demonstrated
quantitatively in Figure 9 and Figure 10, the mean size of the
observed $V$-dropout population is smaller than that predicted based
on a no-evolution extrapolation from the $U$-dropout population with
evidence that the mean surface brightness has evolved too.  Plate
scale is 1.36 arcsec.}
\end{figure}

To examine the rate of evolution in size we repeated the experiment of
cloning the $U$-dropout population to higher redshift but scaled the
sizes by $(1+z)$ (solid lines) without changing the surface
brightnesses.  We also considered the case where the size scaled as
$(1+z)^{-1}$ and the surface brightness varied as $(1+z)$ (dashed
lines).  These lines are presented both in comparison with the
$B$-dropouts (Figure 8) and the $V$-dropouts (Figures 9-10).  As might
be expected, the constant surface-brightness size-evolution model
where size $\propto$ $(1+z)$ overestimates both the sizes and numbers
considerably.  Our other size-evolution model (where sizes decrease as
a function of redshift), however, fares much better, providing a
decent fit to the observations, suggesting that galaxies are
$(1+5)/(1+2.7)\sim1.7$ times larger at $z\sim2.7$ than they are at
$z\sim5$.

It is also instructive to take the angular size distributions of the
$U$, $B$, and $V$ dropouts shown in Figure 7, Figure 8, and Figure 10
(we take the brighter magnitude slices) and represent them in terms of
their intrinsic physical size, assuming they lie at $z\sim2.7$,
$z\sim4$, and $z\sim5$ (Figure 15).  We perform a similar scaling to
the different projections of our $U$-dropout populations shown on
Figures 7, 8, and 10.  It is evident that while the physical sizes of
the $V$-dropout population are significantly smaller than the
$U$-dropouts for all geometries, the extrapolated $U$-dropout
population are also much smaller given the $B$ and $V$ selection
criteria.  Obviously, this latter shift toward smaller intrinsic sizes
must arise from the selection procedure itself and cannot be due to
evolution.  Therefore, the apparent evolution in the size of UV-bright
population can only be partially an issue of evolution.  This
illustrates how important a consideration of selection effects are for
the present analysis.

\begin{figure}
\epsscale{0.95}
\plotone{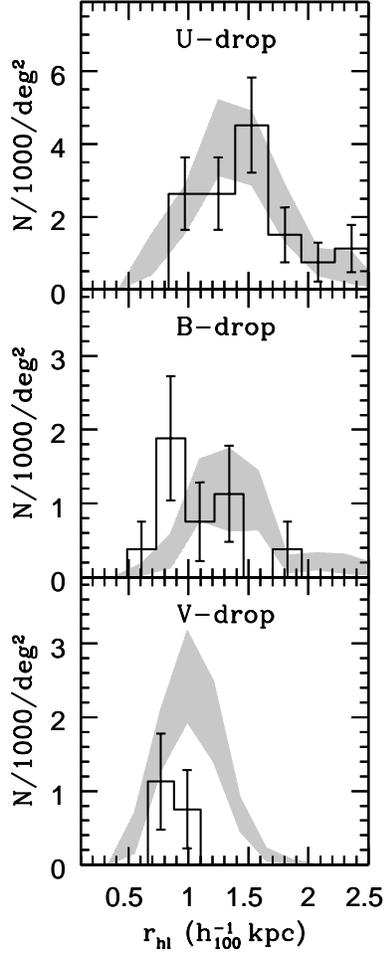}
\caption{A comparison of the physical sizes of the bright $U$-dropouts
($B_{450,AB}<25.5$), $B$-dropouts ($V_{606,AB}<26.5$), and
$V$-dropouts ($I_{814,AB}<27.0$) (histograms) with the sizes recovered
from the extrapolated $U$-dropout population (shaded regions) for the
$\Omega_{M}=0.3$, $\Omega_{\Lambda}=0.7$ geometry.  While the physical
size of the $V$-dropouts is clearly smaller than the $U$-dropouts
(compare the histograms in the top and bottom panels), the projected
$U$-dropouts (via no-evolution) \textit{also} tend to be much smaller
because of the selection effects, so the apparent evolution in the
size of the $UV$-bright population is only partially an issue of
evolution.  This demonstrates how important a consideration of
selection effects can be for measuring evolution.}
\end{figure}

We now attempt to determine the statistical significance of our
finding that galaxies at $z\sim3$ seem to be larger than those at
$z\sim5$.  To this end, we shall suppose that both samples can be
approximated by a normal distribution and we shall test the null
hypothesis that the mean of one sample (our projected $z\sim3$ sample)
is larger than the mean of the other (our observed $z\sim5$
sample.)  Formally, we use the T test:
\begin{equation}
T = \frac{\bar{X}_1 - \bar{X}_2}{\sigma \sqrt{\frac{1}{n_1} + \frac{1}{n_2}}}
\end{equation}
where
\begin{equation}
\sigma = \sqrt{\frac{n_1 S_1 ^2 + n_2 S_2 ^2}{n_1 + n_2 - 2}},
\end{equation}
where $\bar{X}_i$ is the mean for sample $i$, $S_i$ is the variance
for sample $i$, $n_i$ are the number of objects in sample $i$ (e.g.,
Hogg \& Tanis 1993).  We derive T = 1.64 and T = 1.49 for the brighter
($I_{814,AB}<27$) objects in Figures 9 and 10, respectively.  This
works out to a 90\% confidence and 86\% confidence result,
respectively, providing suggestive evidence that $UV$-bright galaxies
are smaller at $z\sim5$ than they are at $z\sim3$.

\subsection{Colour Distributions}

We now look at how the colours of the $B$ and $V$-dropout populations
compare with that expected based upon lower redshift populations to
examine the evolution of intrinsic colours.  Figure 16 compares the
observed color distribution of the $U$, $B$, and $V$ dropout samples
(histogram) with that expected from the $U$-dropout population (shaded
region).  The color distribution at $z\sim4$ and $z\sim5$ seems to
have a very similar shape to that predicted based upon the lower
redshift sample, at $z\sim5$ suggesting that there hasn't been a lot
of evolution in the intrinsic age, metallicity, or dust content of
high redshift galaxies over this redshift interval.

\begin{figure}
\epsscale{0.95}
\plotone{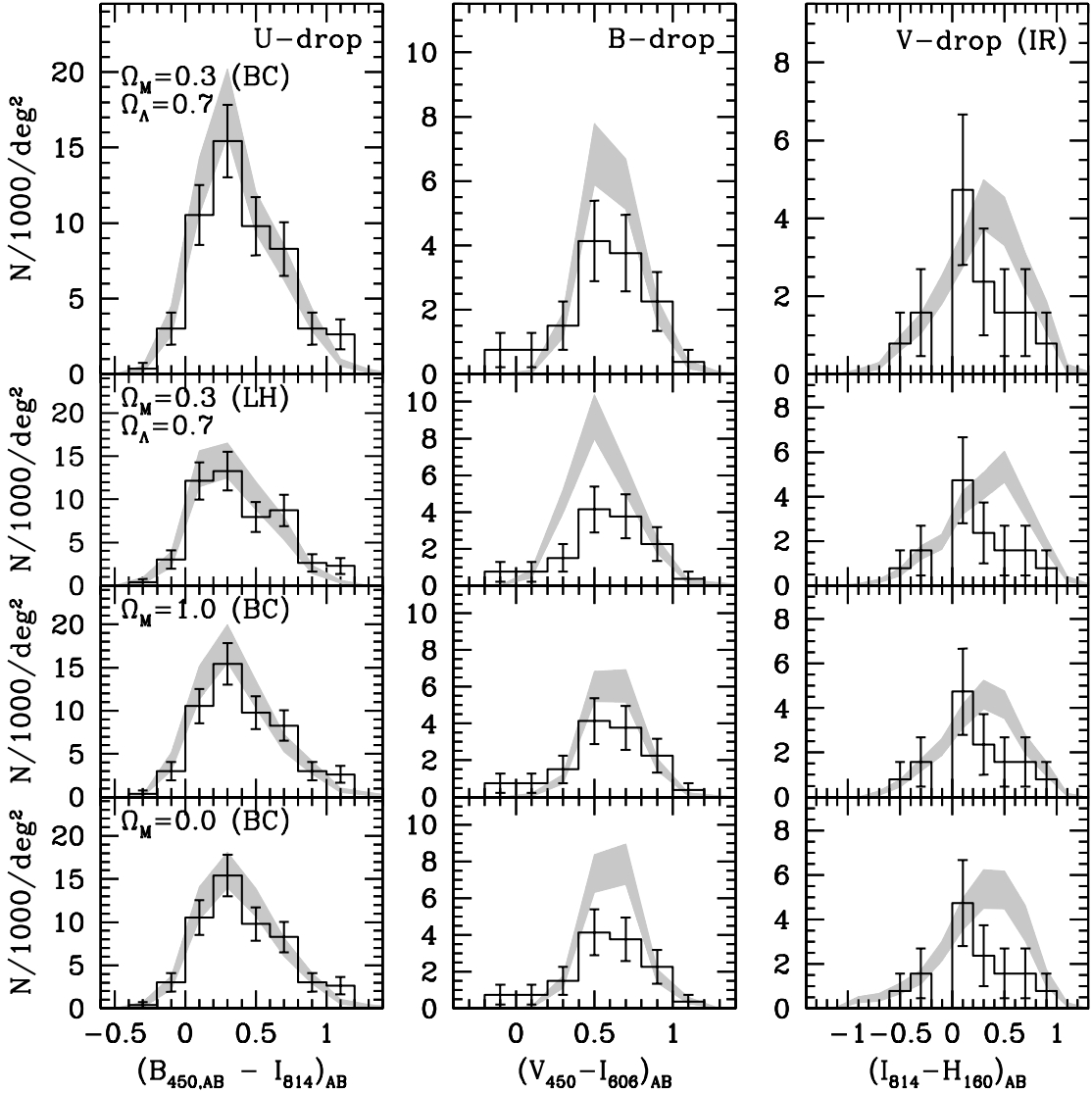}
\caption{A comparison of the simulated (shaded regions) and observed
(histogram) colour distributions for the $U$, $B$, and $V$-dropout
samples.  The shaded regions represent no-evolution expectations based
on our $U$-dropout sample.  We provide the canonical results based on
the $\Omega_{M}=0.3$, $\Omega_{\Lambda}=0.7$ geometry and BC spectral
template set in the top panel.  We also include results using the LH
spectral template set, the $\Omega_{M}=1$ geometry, and $\Omega_{M}=0$
geometry in the lower panels to illustrate possible model
dependencies.  The distribution of intrinsic colors appears to show
minimal evolution from $z\sim5$ to $z\sim3$, suggesting similarly
small changes in the age, metal, or dust-content of this $UV$-bright
population.}
\end{figure}

\subsection{Redshift Distributions}

To illustrate the redshift distributions for the objects in our
dropout samples and to comment on the selection windows used for these
purposes, we compare our estimated redshift distributions (histogram)
with those predicted by extrapolating our $U$-dropout population to
higher redshift (shaded regions).  The redshift distribution for the
$U$, $B$ and $V$-dropouts agree quite well with that expected based
upon the $U$-dropout population.  Note that for both the $U$ and
$B$-dropout samples we observe a downturn at lower redshift than
suggested by Figure 17 of Steidel et al.\ (1999).  This results
because the redder band used in defining the spectral break no longer
has sufficient signal-to-noise at higher redshift to make a strong
constraint on the color.  Consequently, one finds that real objects
make ``inverted-V'' shapes as they track through colour-colour space.
They therefore tend to drop out of the selection window at lower
redshifts than one might naively expect if the object had infinite
signal-to-noise in both passbands defining the spectral break.  This
is illustrated in Figure 18 for 3 objects from our $U$-dropout sample.
The tracks they make in colour-colour space assuming infinite S/N are
also shown (solid line).

\begin{figure}
\epsscale{0.95}
\plotone{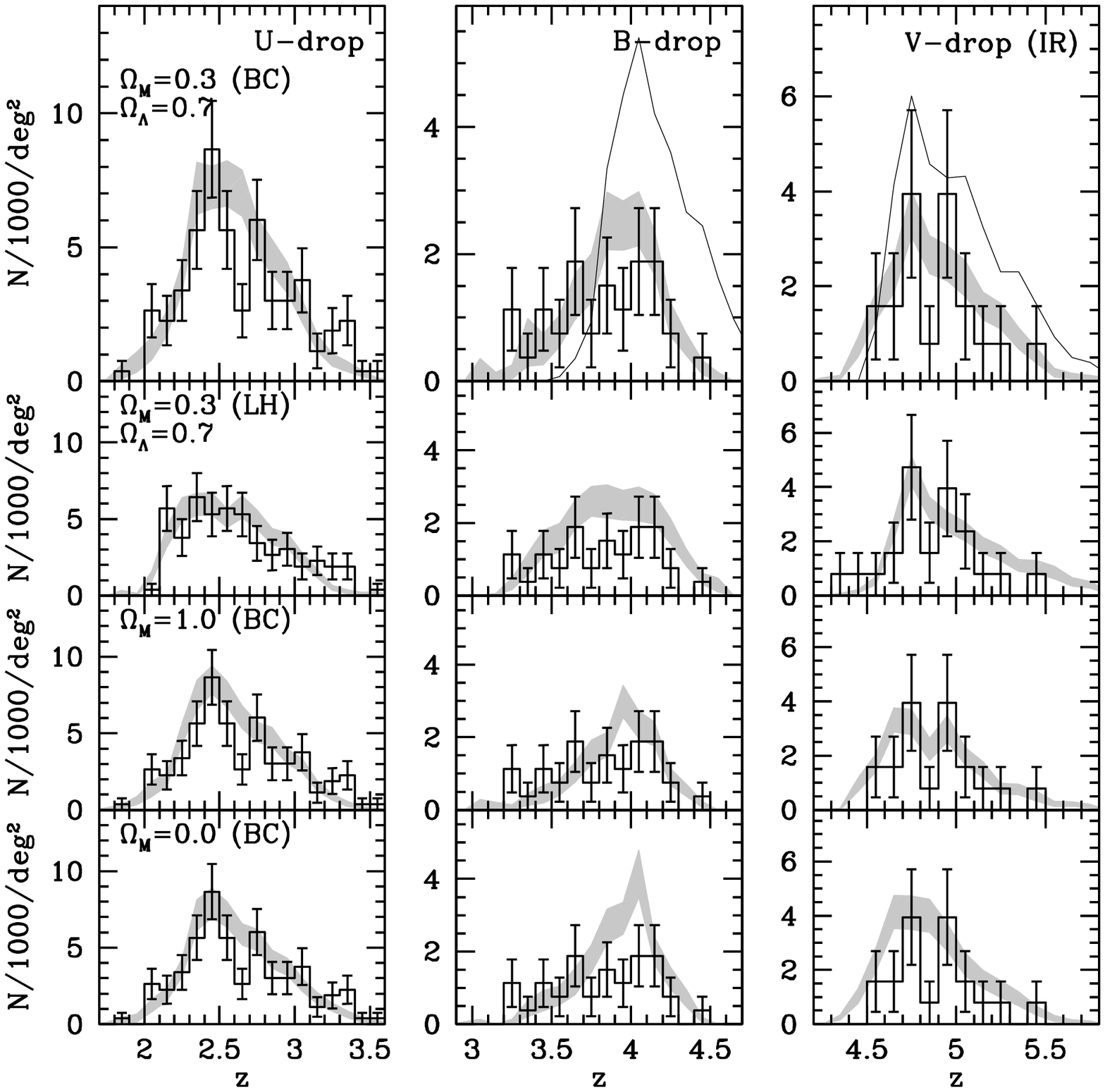}
\caption{Comparison of the simulated (shaded regions) and observed
(histogram) redshift distributions for the $U$, $B$, and $V$-dropout
samples.  The shaded regions represent the no-evolution expectations
based on our $U$-dropout sample.  Note the good agreement between the
predicted $U$-dropout redshift distributions and the observed ones.
The thin solid lines represent the redshift distribution predicted
with no observational errors.  Note that this latter distribution is
higher in both number and redshift than is actually obtained when
observational errors are included, demonstrating the importance of
including such effects.}
\end{figure}

\begin{figure}
\epsscale{0.95}
\plotone{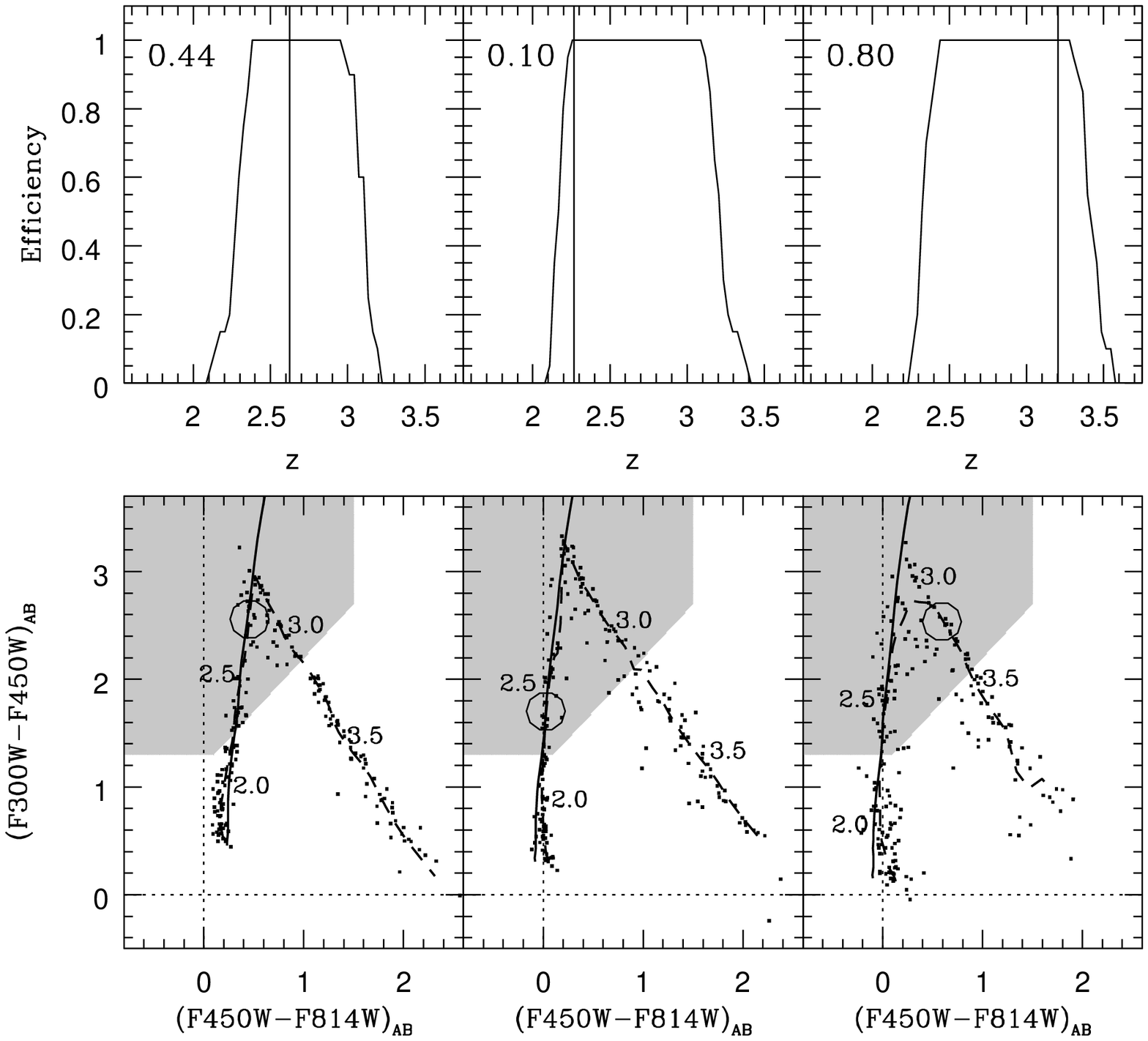}
\caption{The upper panels show typical determinations of the redshift
selection efficiencies $\epsilon(z)$ for three $U$-dropout
galaxies.  Each panel also includes a vertical line at the estimated
redshift and the corresponding $V/V_{max}$ (upper left-hand corner of
each panel).  The lower panels illustrate the Monte-Carlo simulations
we perform to see how the photometry of each object (dots) tracks
across redshift space, specific redshifts being annotated there.  The
solid and dashed lines indicate how the object tracks in color-color
space given infinite and the observed signal-to-noise, respectively.
(The shaded region denotes the $U$-dropout selection window.)  The
actual colors for each object are indicated by the large circles.}
\end{figure}

To illustrate the effect this has on the selection window, we repeat
our extrapolation of the $U$-dropout population to higher redshifts
(\S3.3), but now assume that the object colours simply derive from
their best-fit spectral types.  We include this as thin solid lines in
all panels on Figure 17.  Clearly, many more galaxies are expected to
be selected at high redshift using this method than one finds by
resimulating the object at higher redshift and recovering its
parameters there.  This effect, among other things, may have led
others to overestimate the dropoff in luminosity density from $z\sim3$
to $z\sim4$ based on the HDFs.  This underlines the importance of
doing detailed simulations to understand the selection effects at work
in estimating evolution across a sample.

\subsection{Star Formation History}

We now look at the evolution of the luminosity density.  Typically,
this has been calculated by (1) determining the luminosity function
for a high redshift population, (2) determining the total luminosity
density by integrating along the luminosity function, and (3)
converting the observed luminosity density to a star formation rate
density using some prescription.  Unfortunately, this method suffers
when the implicit set of galaxies selected varies as a function of
redshift.

Here, we proceed as follows.  For the $U$-dropouts, we determine the
luminosity density in the standard way described above, but for the
$B$ and $V$-dropouts, we determine the luminosity densities
differentially, namely, by comparing the observed dropout counts with
that expected from a no-evolution projection of the $U$-dropout
population to higher redshift.  Otherwise stated, we take the
luminosity density of the $U$-dropouts to be
\begin{equation}
L_{UV} (U) = L_{UV} (Obs,U),
\end{equation}
the luminosity density of the $B$-dropouts to be
\begin{equation}
L_{UV} (B) = L_{UV} (Obs,U) (\frac{L_{UV} (Obs,B)}{L_{UV} (Sim,U\mapsto B)}),
\end{equation}
and the luminosity density of the
$V$-dropouts to be
\begin{equation}
L_{UV} (V) = L_{UV} (Obs,U) (\frac{L_{UV} (Obs,V)}{L_{UV} (Sim,U\mapsto V)}),
\end{equation}
where $L_{UV} (Obs,U)$ is the integrated UV luminosity of the observed
$U$ dropouts and where $L_{UV} (Sim,U\mapsto B)$ is the integrated UV
luminosity of the $B$-dropouts recovered from projecting the
$U$-dropout population to higher redshift.  Note that the ratio
$\frac{L_{UV} (Obs,B)}{L_{UV} (Sim,U\mapsto B)}$ is determined by
summing the light found in the observed number counts and comparing
that with the number predicted for our empirical $U$-dropout model.
As remarked in \S3.5, this works out to a measured $UV$-luminosity
density which is 40\% lower at $z\sim4$ than it is at $z\sim2.7$ and
46\% lower at $z\sim5$ than it is at $z\sim2.7$.  On the other hand,
if we had simply determined the luminosity densities at $z\sim4$ and
$z\sim5$ from the luminosity functions estimated there instead of
differentially as we have done here, the shortfall would have been
60\% and 71\%, respectively.

We converted our derived luminosity densities to star formation rate
densities using the relation
\begin{equation}
L_{UV} = \textrm{const}\,\, \textrm{x}\,\, \frac{\textrm{SFR}}{M_{\odot} \textrm{yr}^{-1}} \textrm{ergs}\, \textrm{s}^{-1}\, \textrm{Hz}^{-1}
\end{equation}
where const = ($8.0 \times 10^{27}$, $7.9 \times 10^{27}$) at (1500
$\AA$, 2800 $\AA$) for a Salpeter IMF (Madau et al.\ 1998).  Figure 19
illustrates the present results in the context of other typically
cited determinations of the star formation rate density.

\begin{figure}
\epsscale{0.95}
\plotone{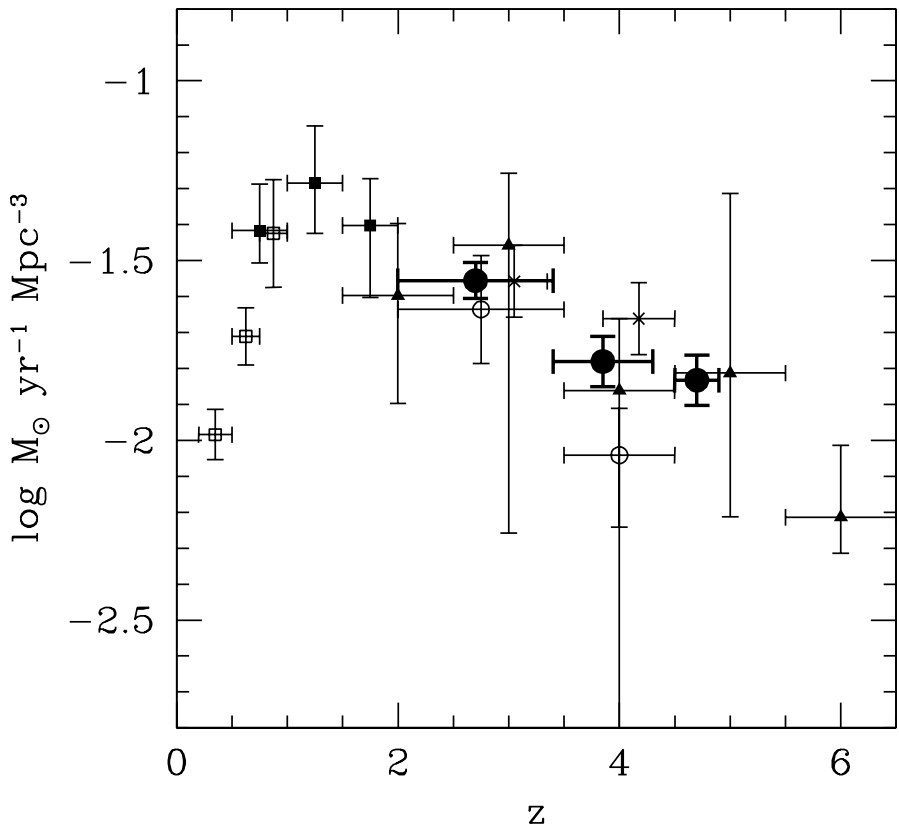}
\caption{A history of the star formation rate density assuming no
extinction correction.  The determinations from this work (large solid
circles) are in fair agreement with the previous high redshift
determinations of Madau et al.\ (1998) (open circles), Steidel et al.\
(1999) (crosses), and Thompson et al.\ (2000) (filled triangles).  The
determinations from Lilly et al.\ (1996) (open squares) and Connolly
et al.\ (1997) (solid squares) are shown for context.  A Salpeter
(1955) IMF is used to convert the luminosity density into a star
formation rate (see, for example, Madau et al.\ 1998).  Values are for
a $\Omega_{M}=0.3$, $\Omega_{\Lambda}=0.7$ geometry and $H_0 = 70\,
\textrm{km/s/Mpc}$.  No correction is made for dust.}
\end{figure}

\section{Possible Dependencies}

All the results we have presented thusfar assume a $\Omega_{M}=0.3$,
$\Omega_{\Lambda}=0.7$ geometry and utilize the BC spectral templates
to perform the $k$-corrections.  Here we repeat the entire analysis we
performed in previous sections, but use different cosmologies and
spectral templates to move the $U$-dropout galaxies through redshift
space.  In particular, we consider the $\Omega_{M}=0$ and
$\Omega_{M}=1$ geometries; and for spectral template sets, we consider
the Leitherer \& Heckman (1995) model for a $10^7$ yr burst and
metallicity $0.2 Z_{\odot}$ with various amounts of dust reddening,
this template set hereafter abbreviated as LH.

In Figures 20-21, we illustrate the effect that different cosmologies
or spectral types have on the size distribution of the $B$ and
$V$-dropouts predicted based upon the $U$-dropouts, respectively.
Note that we smoothed the observed size distribution by different
amounts to mimic the larger effective PSF of objects at $z\sim4-5$.
For both sets of dropouts, similar angular sizes are predicted for the
flat $\Omega_{M}=1$ geometry as for the standard $\Omega_{M}=0.3$,
$\Omega_{L}=0.7$ geometry used earlier in the paper.  For the open
$\Omega_{M}=0$ geometry, however, the angular sizes are predicted to
be smaller and thus in better agreement with the observed size
distribution.  For the $B$-dropout sample, the LH template-set
predictions are $\sim50$\% higher than for the BC templates.  This
results because of the different way the two template sets track
through colour-colour space, one template set having a consistently
higher $B$-dropout selection volume relative to the $U$-dropout
selection volume.  For the $V$-dropouts, however, the LH template set
produces very similar predictions to the BC template set.

\begin{figure}
\epsscale{0.95}
\plotone{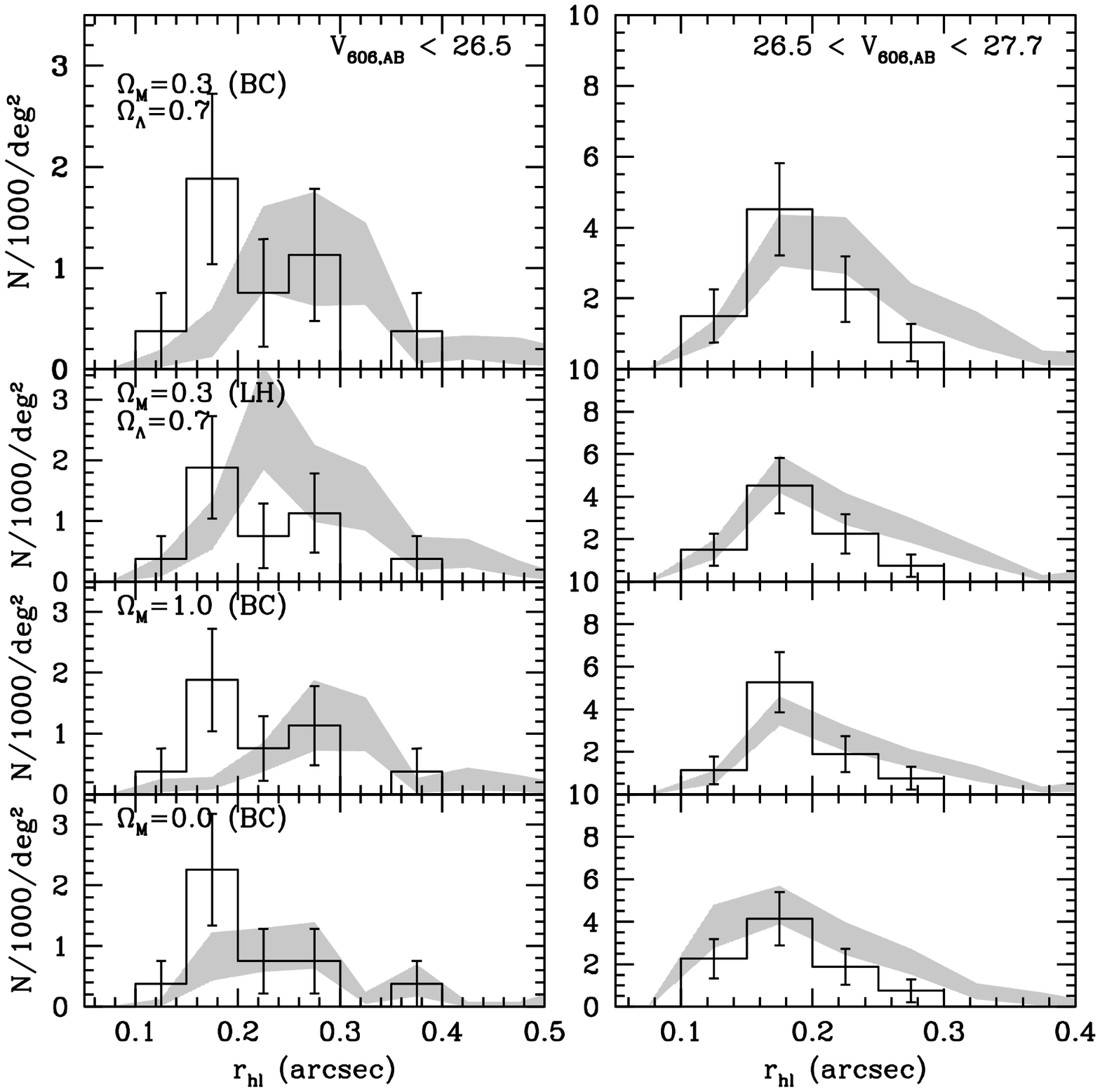}
\caption{Similar to Figure 8, but using different geometries and
spectral template sets to project the $U$-dropout sample to higher
redshifts.}
\end{figure}

\begin{figure}
\epsscale{0.95}
\plotone{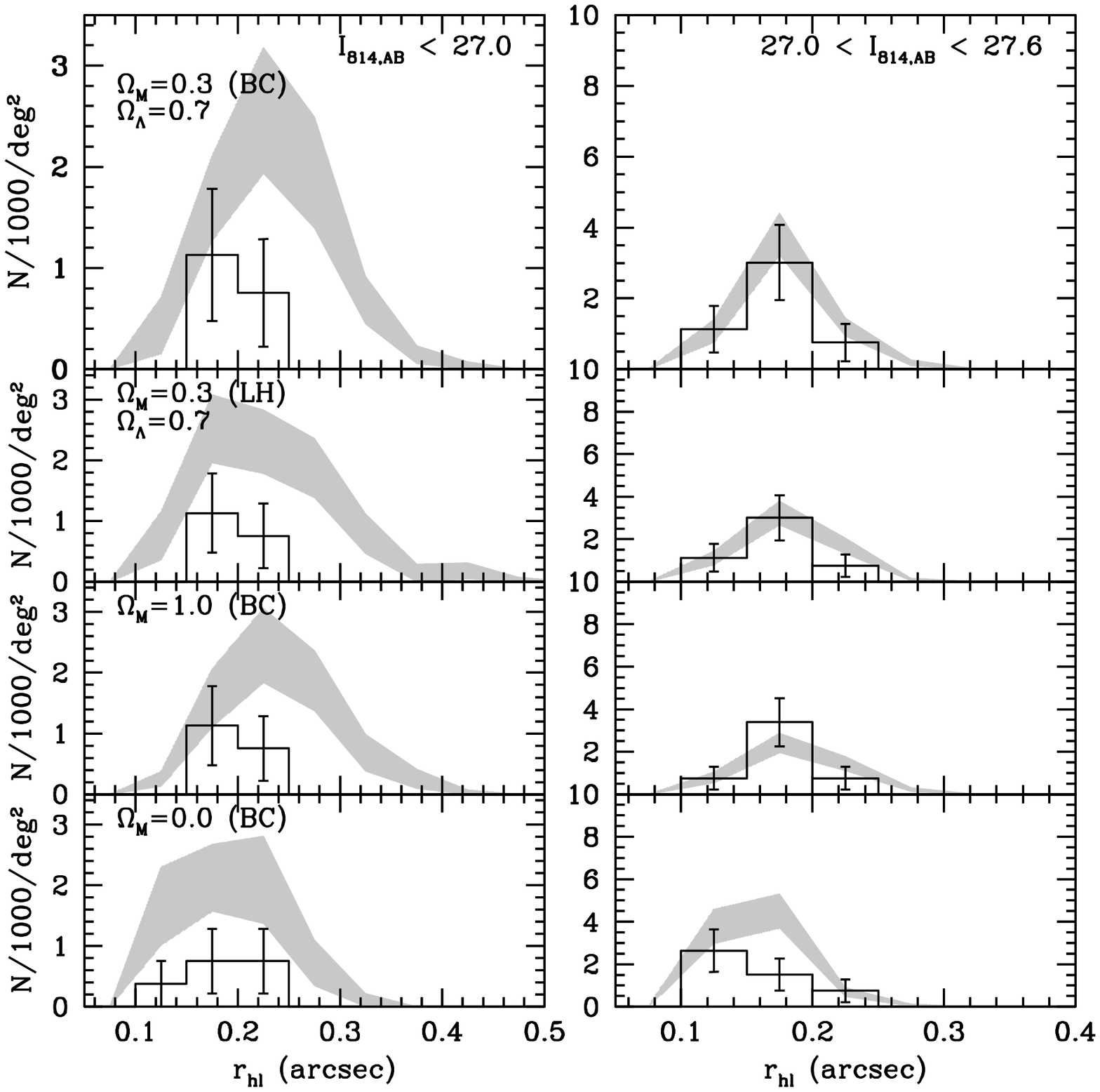}
\caption{Similar to Figure 10, but using different geometries and
spectral template sets to project the $U$-dropout sample to higher
redshifts.}
\end{figure}

In Figures 16-17, we illustrate the effects of cosmology and template
set on both the color and redshift distributions predicted based upon
the $U$-dropouts.  Clearly, there isn't a large dependence on
geometry, but the results do seem to depend a little on the spectral
template set used, calculations involving the Leitherer \& Heckman
(1995) templates being slightly bluer and higher in normalization than
those using the Bruzual \& Charlot (1995) templates.  The color
differences are due to differences in the shape of the spectral
templates--differences which also result in the LH template set having
a consistently higher $B$-dropout selection volume than its
$U$-dropout selection volume.  A systematic study comparing real high
redshift galaxy spectra to these spectral template sets might prove
useful in eliminating these model dependencies.  This, however, is
beyond the scope of the present investigation.

We note here in passing that we checked our results by repeating all
of our calculations assuming the Steidel et al.\ (1999) luminosity
function, the Steidel et al.\ (1999) colour distribution, and real
$U$-dropout profiles.  The predictions we obtained were very similar
(within $10-15$\%) to those obtained by cloning the $U$-dropout
population to higher redshift.  This suggests that the redshift
uncertainties present in our base sample do not have a large effect on
our results.

\section{Discussion}

Up to this point analyses of differential evolution across
high-redshift dropout populations have been restricted to luminosity
functions (Pozzetti et al.\ 1998; Steidel et al.\ 1999), spatial
clustering properties (Steidel et al.\ 1999; Giavalisco et al.\ 1998;
Adelberger et al.\ 1998), and their colours (Steidel et al.\
1999).  Little work has been possible on the differential evolution of
structural parameters.

\subsection{Comparison with Casertano}

It is instructive to compare our results with those of Casertano et
al.\ (2000), who have also determined the number of $U$ and $B$
dropouts in the HDF North and South.  Using nearly identical selection
criteria, Casertano et al.\ (2001) found 68 $U$-dropouts in the HDF
North compared to our 66, and 74 $U$-dropouts in the HDF South
compared to our 76.  For the $B$-dropouts and using a slightly more
conservative selection criteria than ourselves, they found 11 dropouts
in the HDF North compared to our 15, and 18 in the HDF South compared
to our 21.  Besides our use of different photometry, another reason
for the slight differences between our results is that Casertano et
al.\ (2000) do not exclude objects which are likely stars.  Overall,
though, our results agree quite well.

\subsection{Luminosity Functions}

We begin by comparing the luminosity functions we determined with
previous determinations based on the HDFs, in particular those derived
by Pozzetti et al.\ (1998) using the HDF North.  We overplot their
$U$-dropout ($z\sim3$) and $B$-dropout ($z\sim4$) luminosity functions
on Figure 11 using thin solid and dotted lines,
respectively.\footnote{We assume $(M_{1700} - M_{1600})_{AB} =
-0.15$.}  Before discussing a comparison of our derived LFs, let us
remark briefly on their differences.  The Pozzetti et al.\ (1998)
study considers only the HDF North while we include the HDF South.
The Pozzetti et al.\ (1998) study assumes that the selection volume
for all $U$-dropout and $B$-dropout galaxies is uniformly $2<z<3.5$
and $3.5<z<4.5$, respectively, whereas we derive the selection volume
for each galaxy individually, our estimated volume being 30-40\% lower
than the Pozzetti et al.\ (1998) estimate on average.  The Pozzetti et
al.\ (1998) study selects $U$-dropout galaxies in the $V_{606,AB}$
band while we select them in the $B_{450,AB}$ band.  Finally, the
Pozzetti et al.\ (1998) study uses the Kaplan-Meier (Lavalley, Isobe,
\& Feigelson 1992) estimators to determine the number of dropouts.  In
comparing our luminosity functions with those of Pozzetti et al.\
(1998), the bright end of our $U$-dropout LF is higher than the
Pozzetti determination by about 70\%, while the faint ends are more
consistent.  For the $B$-dropout LFs, our bright end also tends to be
a bit higher in normalization while again at the faint end things are
more consistent.  Given the differences in our methodologies, we
believe the differences are largely attributed to the different
assumed selection volumes.

We now move into comparisons with the ground-based results of Steidel
et al.\ (1998), for which there has been broad agreement at $z\sim3$,
but a more controversial discrepancy at $z\sim4$ (Madau et al.\ 1996,
1997; Pozzetti et al.\ 1998; Steidel et al.\ 1999; Casertano 2000).
Accordingly, it is not too surprising that the $U$-dropout luminosity
function we derive at $z\sim3$ (Figure 11) roughly agrees with those
derived from Steidel et al.\ (1999).  We include their $z\sim3$
luminosity function as a set of solid circles and their fit as the
thick solid line, where we take $M_{1700,AB} = -21.18$, $\alpha=-1.6$,
and $\Phi = 0.00198\,\textrm{Mpc}^{-3}$.  Once again, at the bright
end, our $U$-dropout luminosity tends to be a little high, but this is
really quite understandable, especially considering that the scatter
induced by photometric redshift uncertainties increases the number of
galaxies at the bright end and decreases the number of galaxies at the
faint end of the luminosity function.

Also, similar to other results on the HDFs, the $B$-dropout luminosity
function we derive is clearly lower than that of Steidel et al.\
(1998).  We include their $z\sim4$ luminosity function as a set of
open circles and the fit as a thick dotted line, where we take
$M_{1700,AB} = -21.24$, $\alpha=-1.6$, and $\Phi = 0.00154\,
\textrm{Mpc}^{-3}$.\footnote{Fit parameters both here and for the
$z\sim3$ luminosity function are for a $H_0 = 70\,\textrm{km/s/Mpc}$,
$\Omega_{M} = 0.3$, $\Omega_{\Lambda}=0.7$ cosmology.}  While the same
qualifications hold as for the derivation of the $U$-dropout
luminosity function, the $B$-dropout luminosity function is some 60\%
lower in normalization than our $U$-dropout luminosity.  Clearly, this
is somewhat at odds with our finding in \S3.9 that the $z\sim4$
luminosity density is only 40\% lower.

The only obvious way to rationalize these results is to conclude that
there must be a class of objects which is selected at $z\sim3$ by the
$U$-dropout selection criteria and missed by the $B$-dropout criteria.
Typically, the answer would be low luminosity galaxies, but a quick
look at the luminosity functions (Figure 11) shows that this is not
the answer, our $B$-dropout luminosity function probing fainter than
its $U$-dropout equivalent.  Instead, the difference seems to be more
in the color cuts.  Consider both Figure 2 of this paper and Figures
9-10 of Casertano et al.\ (2000).  In both, we find significant
numbers of modestly red galaxies barely excluded from both $B$-dropout
samples (see the $E(B-V)\sim0.3-0.4$ track in Figure 2), objects which
are \textit{included} in the $U$-dropout selection (Figure 1).  Add to
this the likely scenario that several of the lower surface-brightness
$U$-dropout galaxies would tend to be missed in $B$-dropout samples
due to the surface brightness dimming effects and it becomes quite
apparent that there could be large differences in the derived
luminosity functions without significant evolution in the underlying
populations.\footnote{Note that we excluded redder starburst
($E(B-V)>0.35$) galaxies from our $B$-dropout selection function
because of the large number of low-redshift ellipticals which lie very
close to this region in color/color space, thus, making it difficult
to select such objects at high redshifts purely on the basis on their
broadband colors.}  This, in fact, is an excellent illustration of the
difficulties inherent in measuring evolution from the luminosity
function alone and why it is much better to infer evolution
differentially as we have done here.  The reason is simple: the
cloning procedure outlined here automatically corrects for differences
in the selection function in order to estimate evolution.  To make
similar estimates from a set of luminosity functions, one has to be
sure that their selection functions are exactly the same.  This latter
task, quite obviously, is difficult to do cleanly and to the same
surface brightness threshold.  \textit{The preferred approach is
clearly a differential one as used here.}

\subsection{Colour distributions}

Steidel et al.\ (1999) used a model based upon their
observationally-determined luminosity function and distribution of
different dust-reddened spectral types to reproduce their ground-based
$B$-dropout redshift distribution.  As noted in this same work, this
suggests that there hasn't been a dramatic change in the distribution
of UV-bright spectra from $z\sim3$ to $z\sim4$.  We find a similar
result to both $z\sim4$ \textit{and} $z\sim5$, suggesting there has
been minimal evolution in the age, metal, and dust-content content of
$UV$-bright objects from $z\sim3$ to $z\sim5$.  \textit{A priori}
little is known about colour evolution given the huge dependence on
both the content and distribution of dust in these young star-forming
objects.

\subsection{Galaxy Sizes}

At lower redshifts, there has already been a lot of observational work
looking for possible size evolution.  It is still somewhat
controversial whether $L_{*}$ disk galaxies evolve in size from
$z\sim0$ to $z\sim1$ (Lilly et al.\ 1998; Mao, Mo, \& White 1998;
Simard et al.\ 1999; Bouwens \& Silk 2002; Bouwens et al.\ 2002).
However, when low redshift galaxies are compared with fainter, higher
redshift objects ($z>1-2$), there tends to be relatively consistent
agreement that objects become smaller (BBSII; Roche et al.\ 1998; de
Jong \& Lacey 2000; Bouwens \& Silk 2002; Bouwens et al.\ 2003).

In the high redshift interval ($z>2$) little to no observational work
has been done on the differential evolution of galaxy sizes, despite
the availability of high resolution HDF data.  From a theoretical
perspective, one expects the size of virialized objects to scale as
$H(z)^{-2/3}$ for a fixed mass (e.g., Mo, Mao, \& White 1998). This is
simply derived from the scaling of the mean mass density with
redshift, so that galaxies of a fixed mass interior to the virial
radius must be denser at higher redshift.

For all cases but the completely open universe, the matter term
$\Omega_{M}$ dominates at high redshift:
\begin{equation}
H(z) = H_0 \sqrt{\Omega_{\Lambda} + (1 - \Omega_{M} - \Omega_{\Lambda}) 
         (1 + z)^2 + \Omega_{M} (1+z)^3}.
\end{equation}
More simply, for $\Omega_{M}=0.3$, $\Omega_{\Lambda}=0.7$ and
$\Omega_{M}=1$, $H(z) \propto (1+z)^{3/2}$, and so the size of objects
might be expected to scale as $(1+z)^{-1}$ with redshift.  This is
consistent with our observed $(1+z)^{-1}$ scaling. It is far from
clear, however, that an appreciable fraction of the $UV$-bright
objects at high redshift are found in virialized halos with
rotationally cool disks.  Most of them might well be merging gas
clumps within halos that are just beginning to virialize.

\subsection{Luminosity Density}

The luminosity densities we derived from $z\sim2.7$ to $z\sim5$ tend
to be consistent with previous findings (Madau et al.\ 1998; Thompson
et al.\ 2001; Steidel et al.\ 2000) with the possible exception of the
UV density at $z\sim4$ where our estimates are modestly higher than
previous estimates based on the HDFs, but we have already discussed
those differences in \S5.2.  Overall, there is a clear trend towards
lower luminosity densities at high redshift, our $z\sim5$ sample being
lower by some $\sim46$\%.  Interestingly enough, this decrease is very
similar to the expected $(1+z)^{-1}$ fall-off in sizes for our
preferred model from \S3.6: $(1+2.7)/(1+5) \sim 0.62$ or $\sim38$\%.

\subsection{Comparison With Other Methods}

This paper is part of a long-term effort to measure the differential
evolution of galaxies from low redshift to high redshift.  However, it
is by no means the only attempt to make a systematic comparison of
galaxy properties over such a large redshift range.  More than for any
other galaxy property, it has become fashionable to compile the UV
luminosity function (or in its more popular form the star formation
rate density) as a function of redshift (Madau et al.\ 1996, 1998;
Connolly et al.\ 1997; Cowie et al.\ 1999; Yan et al.\ 1999; Steidel
et al.\ 1999; Thompson et al.\ 2001).  Unfortunately, the shear
magnitude of cosmic surface brightness dimming at high redshift makes
the process of comparing high-redshift galaxy populations with low
redshift ones quite difficult.  Specifically, a large population of
disk galaxies could exist at $z\sim3-4$, contribute a significant
amount to the $UV$ luminosity function and cosmic star formation rate
density, and remain entirely undetected given the amount of surface
brightness dimming.

Lanzetta et al.\ (2001) has attempted to address this problem by
examining the cosmic star formation intensity distribution, where a
star formation rate density is assigned to each pixel instead of to
each object.  By looking at the distribution of UV surface
brightnesses instead of the total luminosities, it is relatively
straightforward to compare galaxy populations at a variety of
redshifts and to apply the appropriate cuts in surface
brightness.  Thompson et al.\ (2001) follow Lanzetta et al.\ (2001) in
use of this approach.

While we are encouraged by the attention such approaches give to
important selection effects such as cosmic surface brightness dimming,
we do not favor this approach for a number of reasons.  First,
Lanzetta et al.\ (2001) use photometric redshifts to divide their
galaxies into different redshift samples.  While photometric redshifts
produce very accurate results for a good fraction of objects in the
HDF at low and high redshift, there are still many objects with
redshift degeneracies, i.e., two very different redshifts which are
equally likely, and it is not generally clear that low redshift
objects are not contaminating high redshift samples and high redshift
objects low redshift samples.  This is especially problematic if one
uses a simple maximum likelihood approach since it is functionally
equivalent to using a flat prior in redshift, a prior which
effectively assumes that $L_{*} \propto D_L(z)^2$.  Therefore, one
should not be too surprised that Lanzetta et al.\ (2001) find a
monotonically increasing star formation rate as the likely result of a
few low redshift objects being spuriously assigned to high redshift.
Secondly, Lanzetta et al.\ (2001) do not actually work with the
intrinsic surface brightness distribution of objects at various
redshifts, but instead with that distribution convolved with the
instrumental PSF. For objects whose angular sizes are very similar to
the PSF, this produces strong redshift biases in the star formation
rate intensities derived.  Third, the signal-to-noise of the
individual pixels will generally be much lower in general than the
signal-to-noise associated with the flux of the entire object, and
therefore, using their approach, one would not be able to work with
objects that are as faint as we use in our approach.

The strengths of the present approach center on a procedure where
comparisons between galaxy populations are made directly in terms of
the observables for the lower signal-to-noise population.  The higher
signal-to-noise population is projected onto these observables via the
cloning formalism presented here.  There can be no loss or distortion
of information for the lower signal-to-noise population because it is
not manipulated, and the information in the higher signal-to-noise
population is degraded just enough to match the S/N of the other
population.  In practice, this tends to mean that comparisons between
low and high redshift populations should always be made at high
redshift due to strong cosmic surface brightness dimming.  A corollary
to this is that no conclusions should be drawn about the evolution of
a population of objects across a range of redshifts that cannot be
made from a comparison of their distributions at the high redshift end
of that range.

\section{Summary}

In the paper, we present the formalism and the machinery used to
project one photometrically-selected sample onto another to test for
evolution.  We replicate each object to higher redshift using the
product of volume density, $1/V_{max}$ and the cosmological volume.
Close attention is paid to pixel-by-pixel k-corrections, cosmic
surface brightness dimming, and variations in the PSF.  Objects are
selected and object properties are measured in exactly the same way as
they were in the original sample.  The volume density, $1/V_{avail}$, is
determined by performing similar projections to lower and higher
redshifts.  Simple corrections are also made for both flux and
redshift uncertainties present in the original sample.  Associated
difficulties and challenges are presented and discussed in depth.

With this machinery, we have addressed the evolution of high redshift
galaxies in the HDF North and South by replicating the $U$ dropout
sample to higher redshift for a fully empirical no-evolution
comparison with the $U$, $B$, and $V$ dropout samples from the same
fields (our cloning procedure).  We find that
\begin{itemize}
\item{The UV luminosity density as inferred from the total integrated
luminosity in the $U$, $B$, and $V$ dropouts is $\sim$46\% lower at
$z\sim5$ than it is at $z\sim2.7$ (an increase of 1.85$\times$ from
$z\sim5$ to $z\sim2.7$).}
\item{We note that the evolution in the UV luminosity density inferred
using our cloning approach is somewhat less than one obtains from a
comparison of the luminosity functions themselves, an increase of
1.7$\times$ from $z\sim4$ to $z\sim3$ using the former method versus
an increase of 2.7$\times$ over the same redshift interval using the
latter method.  As argued in \S5.2, our cloning approach should give
the more accurate results, since it automatically corrects for
differences in the selection functions, our $B$-dropout selection
being less sensitive to redder, lower surface-brightness galaxies.}
\item{We use our empirical cloning procedure to derive a selection
volume for the $B$-dropouts and find a value that is $\sim30$\% lower
than one would obtain by simply redshifting SED templates through the
filter bandpasses.  We also find a slightly lower mean redshift,
$z\sim3.85$, than was used in previous studies (Madau et al.\ 1996).
For both this point and the former, we infer a less severe falloff in
the $UV$ luminosity density from $z\sim2.7$ to $z\sim4$ than others
have reported using the HDFs.}
\item{For both flat ($\Omega_M = 1$) and Lambda-dominated
($\Omega_M=0.3$, $\Omega_{\Lambda} = 0.7$) universes, the mean object
size increases by about $\sim$70\% from $z\sim5$ to $z\sim2.7$
consistent with a $(1+z)^{-1}$ scaling of size with redshift, i.e.,
objects are 1.7$\times$ larger at $z\sim2.7$.  For an open universe,
no significant size evolution is required to occur over this redshift
range, due to the larger change in the angular diameter-distance
relation.}
\item{The distribution of intrinsic colors, as inferred from a
comparison of the $V$-dropout $I_{814}-H_{160}$ colors with that
expected based upon the $U$-dropouts, exhibits minimal evolution over
the redshift interval $z\sim5$ to $z\sim3$.  Similar consistency is
found between the $B$-dropout $V_{606}-I_{814}$ colors and that
expected based upon the $U$-dropouts.  This suggests that there has
been little change in the intrinsic distribution of ages, metallicity,
and dust-content for $UV$-bright objects over this redshift interval.}
\end{itemize}

We have presented strong evidence pointing toward a general increase
in the mean size of objects from $z\sim5$ to $z\sim2.7$ using the $U$,
$B$, and $V$ dropout samples.  While a number of studies already point
toward a significant decrease in size from $z\sim0.5$ to $z>1$ (BBSII;
Roche et al.\ 1998; de Jong \& Lacey 2000; Bouwens \& Silk 2002;
Bouwens et al.\ 2003), it would be interesting to use the same
machinery described here to try to quantify how lower redshift
galaxies--i.e. Balmer break galaxies at $z\sim1$ or even lower
redshift galaxies--fit into this size evolution trend illustrated
here.

The machinery presented in this paper is of generic utility beyond the
task for which it was employed here.  It is useful for measuring
evolution across any purely photometrically-selected astrophysical
samples.  Obvious topical applications include measuring the space
density evolution of barred galaxies, measuring the space density
evolution of elliptical galaxies, the space density and size evolution
of disk galaxies and evaluating the rather large lensing corrections
in ground-based weak lensing measurements.

\acknowledgements

RJB would like to expressly thank Narciso Ben{\'i}tez, Emmanuel
Bertin, Fred Courbin, Harry Ferguson, Andy Fruchter, and Dan Magee for
input helpful for the development of our cloning software.  We would
also like to thank Kurt Adelberger, Narciso Ben{\'i}tez, Daniela
Calzetti, Mark Dickinson, Harry Ferguson, Claus Leitherer, and Chuck
Steidel for valuable conversations related to this project and Mark
Dickinson for generously providing us with his fully reduced HDF North
NICMOS data.  RJB gratefully thanks the European Southern Observatory
for hosting him while a large part of the present work was done.  RJB
acknowledges support from an NSF graduate fellowship, and TJB
acknowledges the NASA grant GO-05993.01-94A.  GDI and RJB acknowledge
the support of NASA grant NAG5-7697.

{}

\appendix

\section{Cloning Procedure I (Object Definition)}

\subsection{Sample Selection}

The first step in cloning a galaxy sample is to select the sample
galaxies themselves, a process which involves both object detection
and photometry.  We perform object detection by adding relevant images
together in quadrature--here the F300W, F450W, F606W, and F814W
images--degrading their PSFs to match the broadest PSF and weighting
them by the reciprocal of the noise to produce so-called $\chi^2$
images following the recipe given by Szalay, Connolly, \& Szokoly
(1999).  We then smooth the detection image with a kernel and look for
$5 \sigma$ peaks.  We take our smoothing kernel to be a Gaussian with
a width equal to the sum in quadrature of $\sigma_{\chi^2}$ and the
pixel size (0.04 arcseconds) where $\sigma_{\chi^2}$ is the sigma of
the Gaussian which best fits the effective PSF of the detection image.

We then do photometry on the detected objects.  No single aperture
perfectly balances the somewhat contradictory aims of measuring the
total object flux (out into the wings) and maximizing the
signal-to-noise ratio for this flux.  Roughly speaking, the smaller
the aperture, the higher the signal-to-noise, but the more flux one
misses.  Conversely, the larger the aperture, the more flux one picks
up, but the lower the signal-to-noise.  We, therefore, find it useful
to use two apertures, both a small and large one.  We use the small
apertures to derive high signal-to-noise estimates of the flux in each
passband, and we use the large apertures to estimate the amount of
flux missed in the small apertures.  We apply the same small-to-large
aperture correction for all passbands so that any flux uncertainties
in the wings of an object are not folded into the estimated color.  We
derive this small-to-large aperture correction from the detection
image.  Both our small and large apertures enclose elliptical regions
sized to have major and minor axes equal to some multiple $k$ of those
same moments for the object (Kron 1980).  We take this multiple $k$,
elsewhere called the Kron Parameter, to be equal to 1 and 3.5 for our
small and large apertures, respectively.  We measure magnitudes in
these adaptive apertures using the MAG\_AUTO parameter available in
processing images with SExtractor.

We assign redshifts to objects either by explicitly matching them with
catalogues of spectroscopic redshifts where available, or by
estimating the redshift from the photometry.  Fortunately, for the
present sample, a sizable fraction of the brighter galaxies have
measured redshifts, e.g., Cohen et al.\ (2000), and so not only have
we been able to use spectroscopic redshifts for many of the objects in
our sample, but we have been able to test the reliability of our
photometric redshift estimates. A convenient compilation of most of
these redshifts can be found in Papovich et al.\ (2001).

The likelihood of a particular redshift $z$ and spectral template
$T_{E(B-V)}$ can be expressed as follows:
\begin{equation}
P(T_{E(B-V)}) e^{-\chi^2 (z,T_{E(B-V)})}
\end{equation}
where
\begin{equation}
\chi^2 (z,T_{E(B-V)}) = \sum _i \left (
\frac{f_i - f_{i,mod} (z,T_{E(B-V)})}{\sigma_i} \right )^2
\end{equation}
where $f_i$ is the flux in the band $i$, where $f_{i,mod}
(z,T_{E(B-V)})$ is the flux of some model SED $T_{E(B-V)}$ at redshift
$z$ in band $i$, where $T_{E(B-V)}$ are the Bruzual \& Charlot (1995)
dust-reddened templates that we described above and where
$P(T_{E(B-V)})$ is the prior reflecting the intrinsic distribution of
templates, which we take to be as follows:
\begin{equation}
P(T_{E(B-V)}) = \left\{
\begin{array}{ll} 2,& E(B-V) < -0.15,\\
24,& -0.15 < E(B-V) < 0.1,\\
38,& -0.1 < E(B-V) < 0.05,\\
39,& -0.05 < E(B-V) < 0.0,\\
120,& 0.0 < E(B-V) < 0.05,\\
110,& 0.05 < E(B-V) < 0.1,\\
185,& 0.1 < E(B-V) < 0.15,\\
128,& 0.15 < E(B-V) < 0.2,\\
140,& 0.2 < E(B-V) < 0.25,\\
110,& 0.25 < E(B-V) < 0.3,\\
45,& 0.3 < E(B-V) < 0.35,\\
30,& 0.35 < E(B-V) < 0.4,\\
13,& 0.4 < E(B-V).\\
\end{array}
\right.
\end{equation}
The above distribution is intended to be identical to the intrinsic
color distribution given in Steidel et al. (1999) (see Figure 6 from
that paper).  While negative values of $E(B-V)$ are clearly
unphysical, they are used for similarity with Steidel et al.\ (1999)
since they provide a convenient way of representing templates bluer
than the base spectral template described above.

Ideally, one would include Monte-Carlo realizations of the Lyman-alpha
forest in our calculation of the expected broadband colours instead of
simply the mean extinction as given by Madau (1995).  Fortunately,
deviations from this mean extinction over broad passbands tend to be
rather small (Bershady, Charlton, \& Geoffroy 1999), the only possible
exception being objects near the epoch of reionization.

\subsection{Object Extent}

Using the image segmentation maps produced by SExtractor from the
$\chi^2$ images, we determine the two-dimensional extent of each
galaxy on the image.  We then enlarge this region to include all
pixels within 4 half-light radii and which do not belong to other
objects.  For pixels which fall within 4 half-light radii of two
different galaxies, priority is given to the galaxy with the higher
value of $(\chi^2\,\textrm{flux})/(r_{hl}^2)$.  From this
two-dimensional pixel-map we make a two-dimensional galaxy template,
the pixels not belonging to the galaxy being filled with noise and
smoothed according to the properties of the field in question.

\subsection{Pixel-by-pixel SED Representation}

Clearly, when replicating an object to different passbands or to
different redshifts, one needs a method for determining how it will
appear at arbitrary rest-frame wavelengths.  Obviously, when the
rest-frame wavelength corresponds with one of the template images, one
would like to use the template image itself, this being the most
model-independent solution.  Similarly, when the rest-frame wavelength
is in between that sampled by two template images, one would like some
way to interpolate between them.  Finally, when the rest-frame
wavelength is beyond the range of the template images, one would like
some suitable way of performing an extrapolation.  In order to
simultaneously accomplish all these aims, we determine the best-fit
SED templates and bolometric surface brightnesses (two degrees of
freedom) for each pixel.  This results in an empirical model image,
and we can use it to resimulate an object at arbitrary redshift.  We
also keep track of the difference images between the observations and
best-fit model images to insure that we can resimulate an object
exactly as it appeared on the original images and with exactly the
\textit{observed} SED.

Before the pixel-by-pixel fits are performed, however, all relevant
images must use the same PSF.  For any given sample, a given passband
is chosen to have the representative PSF which we henceforth call the
representative image.  This is typically the highest-resolution
passband which still has a reasonable pixel-by-pixel S/N and is
somewhat subjectively chosen when the original sample is defined.  A
simple consideration of the diffraction limit ($\sim\lambda/D$) might
lead us to use the bluest passband (where there is still flux) as the
representative image for the present study.  Unfortunately, due to
WFPC2's severe undersampling problems and additional smoothing brought
about by the pixel response function, there isn't a very large
difference in the effective PSF across the different HDF images (less
than 7\% in the FWHM on the few stars we measured).  Because of these
marginal differences, we take the reddest image, $I_{814}$, as
representative.

We determine the best-fit bolometric surface brightness $I_{i,j}$ and
SED template $SED_{i,j}$ by minimizing $\chi ^2$:
\begin{equation}
\chi^2 = \sum _{X} 
\left (\frac{I_{i,j} g(SED_{i,j},X,z) - f_{i,j} ^ {X}}{\sigma_{i,j} ^{X}} \right)^2 
\end{equation}
where $g(SED_{i,j},X)$ is the flux of template $SED_{i,j}$ in band $X$
at redshift $z$ and where $f_{i,j} ^{X}$ and $\sigma_{i,j} ^{X}$ are
the flux and uncertainty in the $X$-band flux at pixel position
($i,j$), respectively.  For each band, we store the differences
between the observed and best-fit images $I_{i,j} g(SED_{i,j},X,z)$:
\begin{equation}
\Delta f_{i,j} ^X = f_{i,j} ^ {X} - I_{i,j} g(SED_{i,j},X,z) 
\end{equation}
Hence, for each object, the total number of images we store is equal
to two plus the number of images used in deriving the clone.  With
this information, we can simulate galaxies exactly and with the same
noise properties as they had on the original images.  Here we used our
Bruzual \& Charlot (1995) dust-reddened starburst templates.

Obviously, for pixels with low signal-to-noise, there are few
constraints on which SED template to use.  Fortunately, in these
cases, the most likely values for the bolometric surface brightness
are going to be smaller than the noise on any one image, so the
uncertainties in the best SED template will be mitigated by the small
size of the derived bolometric surface brightness.  Obviously, in the
case that one extrapolates the flux well beyond the observed
wavelength baseline (for a $z=3$ galaxy observed in the HDF, this
baseline would be between 875 $\textrm{\AA}$ and 2000 $\textrm{\AA}$),
one's results could quickly become inaccurate and most certainly would
be model-dependent.

\section{Cloning Procedure II (Object Replication)}

Here we detail the basic procedure used to replicate objects to
different redshifts and passbands, the basic steps of which are
illustrated in Figure 22.  

\begin{figure}
\plotone{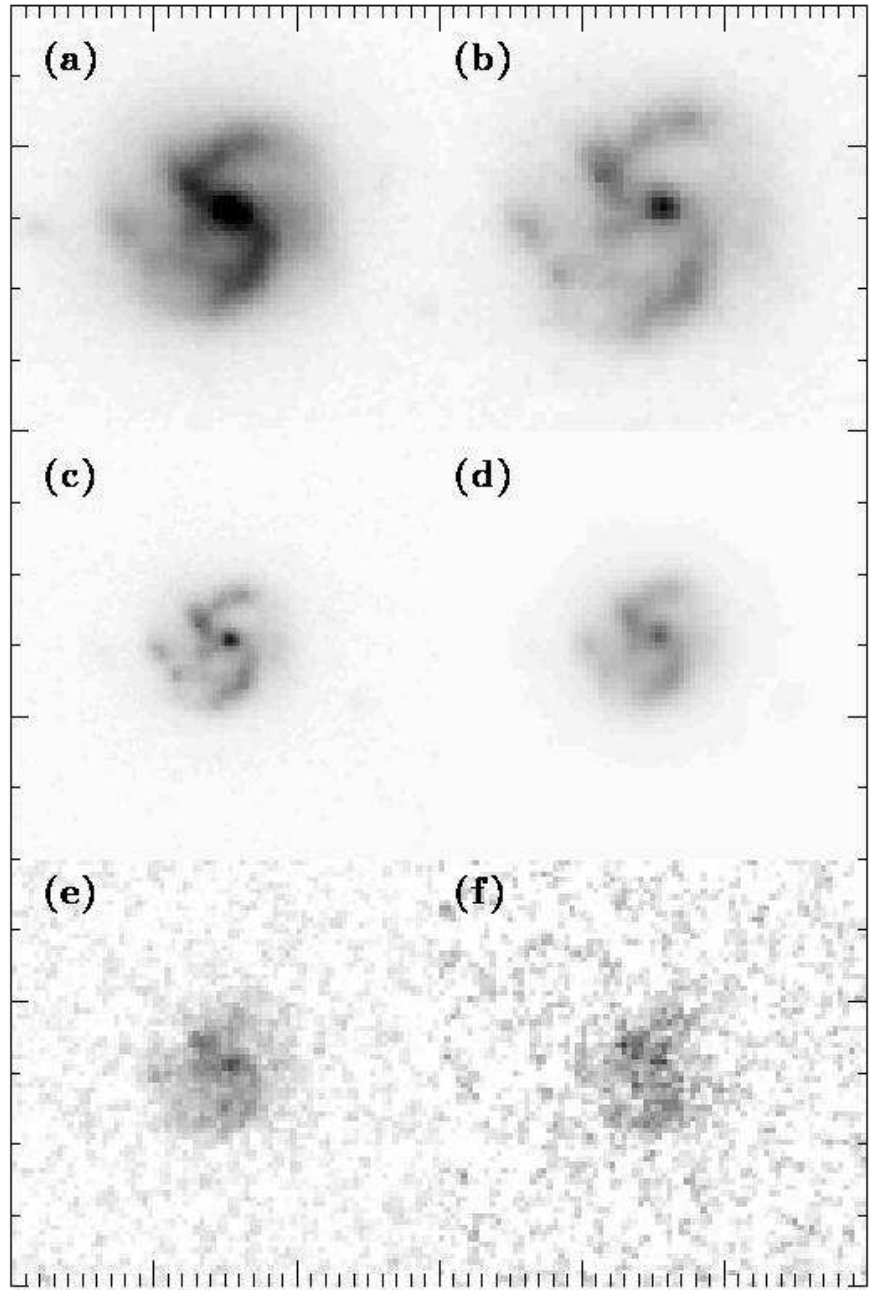}
\caption{Steps involved in projecting a galaxy to higher redshift.
The original image of a galaxy at z=0.47, taken from the HDF (a) is
k-corrected to z=1.5 (b), reduced in angular size to match this
increased redshift (c), resampled with additional PSF smoothing to
compensate for its smaller size (d), and covered with additional noise
to match its fainter magnitude (e).  Panel (f) matches a 25ks WFPC2
exposure ($\sim0.7^m$ shallower than the HDFs).}
\end{figure}

\subsection{Pixel-by-pixel K-corrections}

As discussed above, an important first step in determining the
appearance of an object at different redshifts and for different
passbands is to calculate its surface brightness distribution at an
arbitrary rest-frame wavelength.  To this end, we construct a
pixel-by-pixel template for an object using the formalism discussed
above.  For an object observed in the $Y$ band at redshift $z$, the
template is equal to the sum of a model term and a correction term
\begin{equation}
I_{i,j} g(SED_{i,j},Y,z)
+ \Sigma _{X=X_L} ^{X_H}
\left( \frac{|\lambda ^Y - \lambda^X
\left( \frac{1+z}{1+z_{obs}} \right) |}{\lambda^{X_H} \left( \frac{1+z}{1+z_{obs}} \right)-\lambda^{X_L} \left( \frac{1+z}{1+z_{obs}} \right)} 
\right) 
\left( \frac{1+z}{1+z_{obs}} \right ) ^4 \Delta f _{i,j} ^X
\end{equation}
where $\lambda ^Y$ is the central wavelength of band $Y$.  The band
$X$ in the summation above ranges over the passbands whose redshifted
central wavelength $\lambda^{X} \left( \frac{1+z}{1+z_{obs}} \right)$
most closely straddles the central wavelength of the observational band
$Y$, $\lambda ^{Y}$--where $z_{obs}$ is the redshift of the original
object on which the templates are based.  Hence, for a galaxy
originally observed in the $U_{300}$, $B_{450}$, $V_{606}$, $I_{814}$
passbands at $z\sim3$ and resimulated at $z\sim4$ in the $V_{606}$
band, $X$ would include both the $B_{450}$ and $V_{606}$ bands,
$\lambda^{B_{450}} \left( \frac{1+z}{1+z_{obs}} \right) \sim (450
\,\textrm{nm})(5/4) \sim 563\,\textrm{nm}$ and $\lambda^{V_{606}}
\left( \frac{1+z}{1+z_{obs}} \right) \sim (606\,\textrm{nm})(5/4) \sim
758\,\textrm{nm}$, straddling the central wavelength of $V_{606}$,
$\sim606\, \textrm{nm}$.  Obviously, when $\lambda ^Y$ is outside the
range of the redshifted templates, there will only be one passband $X$
in the above summation.

For the observed redshifts and passbands, Eq. (B1) gives exactly the
same images as found in the processed data.  Only to the extent that a
simulated passband fails to line up with one of the redshifted
passbands are the derived templates based upon the best-fit SED
templates.  Note that this is in contrast to and an improvement over
the procedure used in BBSI (where the $\Delta f_{i,j}$ terms were
dropped) as it depends less on the details of the SED templates used.

\subsection{Geometric Corrections}

We then orient the k-corrected galaxy template on our simulated image,
scaling the template pixel size according to the canonical angular
size-distance relationship, 
\begin{equation}
d_{temp} = d_{temp,obs} \frac{D_A(z)}{D_A (z_{obs})}.
\end{equation}
where $d_{temp}$ is the pixel-size of the projected template,
$d_{temp,obs}$ is the pixel-size of the template in the original
image, and $D_A$ is the angular-size distance relationship.  We lay
the templates down on the simulated images with a variety of pixel
centers and rotation angles.

\subsection{Correcting the PSF}

After the galaxy template has been laid down on the image it is
standard to convolve it with the instrumental PSF, henceforth called
\textbf{PSF-new}.  Unfortunately, the original galaxy template has
already been implicitly convolved with a PSF (henceforth called
\textbf{PSF-implicit-unscaled}) by virtue of its being drawn from some
set of observations.  So, to determine the PSF that still needs to be
convolved with the image (hereafter called \textbf{PSF-corr}), one
needs to deconvolve \textbf{PSF-implicit-unscaled} scaled according to
the angular distance relationship (hereafter, called
\textbf{PSF-implicit-scaled}) from \textbf{PSF-new}:
\begin{equation}
\textbf{PSF-corr} * \textbf{PSF-implicit-scaled} =
\textbf{PSF-new}
\end{equation}
where $*$ indicates a convolution.  Note that by performing the
deconvolution on high S/N PSFs rather on the images themselves, we
avoid performing any deconvolutions on the observational images
themselves and therefore degrading the signal-to-noise.

We employ 50 iterations of the Lucy-Hook algorithm (Lucy 1974) to
accomplish this.  In each iteration, we compute
\begin{equation}
f(n+1) = f(n) \textbf{PSF-implicit-scaled} * \left[
\frac{\textbf{PSF-new}}{f(n) * \textbf{PSF-implicit-scaled}}
\right ]
\end{equation}
For our initial guess, we take the best-fit $\sigma$'s from fits of
our PSFs to Gaussians, and then scale the width of \textbf{PSF-new} by
$\sqrt{1-(\sigma_{implicit-scaled} / \sigma_{new})^2}$ where
$\sigma_{implicit-scaled}$ and $\sigma_{new}$ are the $\sigma$'s of
the best-fit Gaussian to \textbf{PSF-implicit-scaled} and
\textbf{PSF-new}, respectively.  To reduce the number of
deconvolutions needed we tabulate results for each pair of
\textbf{PSF-new} and \textbf{PSF-implicit-scaled} for 10 different
values of $\sigma_{implicit-scaled}/\sigma_{new}$ varying from 0 to
1.  We make a simple interpolation between the results.  Finally, we
convolve \textbf{PSF-corr} with the simulated image.

\subsection{Adding Noise}

At the end of the process we add noise to each pixel.  This is not
completely trivial.  Since the galaxy templates already contain noise,
the simulated images generated from these templates will also contain
noise.  It is, therefore, necessary when simulating images in B.1 to
keep track of how much noise has been added to the individual pixels
of the simulated image so we can add the remainder at this step.  Note
also that before adding this additional noise to the image it is
smoothed to reflect the correlation properties of the noise for the
images in question, here the HDF drizzled images.  The kernel for the
drizzled WFPC2 images is given in Williams et al.\ (1996).

Now, we proceed to describe the task of estimating noise on the galaxy
templates derived from Eq. (B1).  While one might suppose this task to
be relatively straightforward, it is a little more subtle than one
might imagine.  This subtlety owes itself to the fact that simulated
postage stamps are the sum of two terms, a k-corrected model profile
$I_{i,j}$ and at least one correction image $\Delta f_{i,j}$, both of
which contain noise (see Eq. (B1)).  Recall that the model profile
$I_{i,j}$ is a fit to the pixel-by-pixel values for a set of images,
and so for fits on the outer portion of the postage stamp where the
signal is dominated by the noise, both the SED $SED_{i,j}$ and the
profile itself $I_{i,j}$ will contain noise, and that once the postage
stamp is transposed to another redshift, as per Eq. (B1), this noise
will also be present, but its magnitude will depend upon the
distribution of best-fit SEDs used to represent that noise and the
size of the k-corrections made to each of those best-fit SEDs.

Due to the aforementioned subtleties, perhaps the best way of
estimating noise on a galaxy template calculated from Eq. (B1) is in
exactly the same way one measures noise off an astronomical image.
Unfortunately, not all templates are very large and some are largely
covered by an object.  A simple workaround was to simulate a set of 15
x 15 noise images to process alongside the postage stamps derived from
the raw data.  In other words, just as for pixels of an image
containing a real object, we determine the best-fit model profiles $I$
and SED templates $SED$ for the noise patch (and correction image(s)
$\Delta f$) and then transpose it to the object redshift using
Eq. (B1).  Of course, at this point, it is quite straightforward to
determine the noise properties of the transposed noise patch.

\subsection{Analysis of Simulated Images}

Having simulated postage stamps using the steps outlined in Appendix
B.1-B.4, we use exactly the same computer code to analyse the
simulated postage stamps as we use on the original images, e.g., the
procedure outlined in Appendix A.1.  In contrast to our previous work
(BBSI) where we generated large galaxy images and analysed them, we
prefer to recover all the properties from small postage stamps on
which the object has been added.  This speeds up the calculations
considerably and allows us to look at the selection effects related to
object redshifting independent of those effects related to object
overlap.  

We do not perform background subtraction on our simulated images
because a nonnegligible amount of the template flux itself is
typically included in the background determination and thereby biases
the magnitudes recovered.  Embedding the object in the middle of a
much larger image effectively removes this bias, but makes the entire
process much more computationally expensive.  We therefore exclude the
background-subtraction step altogether cognizant of the fact that our
simulations effectively ignore a small source of scatter resulting
from uncertainties in background subtraction step (estimated to be up
to a $\sim10\%$ effect depending on the object and surface density of
its neighbors).

\section{Volume Density}

The next step is to estimate the volume density of each object in the
original sample so that we know how frequently to replicate it in
simulated samples.  We take the volume density of each object to be
equal to
\begin{equation}
n = \frac{1}{V_{avail}}.
\end{equation}
where $V_{avail}$ is the expected effective volume in which the object
would fall into the selection sample.  Formally, the volume available
$V_{avail}$ is taken to be
\begin{equation}
V_{avail} = \Omega
\int _{z=0} ^{z=\infty} \epsilon (z) \frac{dV}{dz d\Omega} dz
\end{equation}
where $\epsilon (z)$ is the efficiency of detection at each redshift
and $\Omega$ is the selection area in steradians.  The efficiency
$\epsilon(z)$ accounts for the extent to which photometric scatter
places a galaxy inside or outside the sample.  We calculate $\epsilon
(z)$ by Monte-Carlo resimulating and remeasuring the galaxy at
different orientation angles using the procedures laid out in Appendix
B.1-B.5.  We add magnitude scatter to the results using the expected
Poissonian and Gaussian noise.  Ideally, we would add each galaxy to
different parts of a frame with all foreground and background objects
present to account for the effects of overlap on object detection and
parameter extraction (important for perhaps 2-3\% of galaxies) but,
due to the high computational cost of recovering object parameters for
all sample galaxies over a range of both redshift and galaxy
environment, we chose not to do this.  In Figure 18, we illustrate
Monte-Carlo realizations of the photometry for 3 sample $U$-dropouts
as a function of redshift.

\section{Challenges}

\subsection{Uncertainties in Empirical Clone Models}

While the procedure of mapping one sample of galaxies to another set
of redshifts for comparison with another sample is extremely model
independent in principle, uncertainties both in the pixel-by-pixel
fluxes and the photometric/spectral redshift necessitate the
introduction of small model dependencies into the procedure.  Even
pixel-by-pixel variations in the amount of intervening hydrogen
introduce uncertainties, though studies (e.g. Bershady et al.\ 1999)
have shown that the variation is typically small.  For some samples
such as a bright ($I<23$) HDF sample with spectroscopic redshifts,
these uncertainties are minimal and the results therefore largely
model-independent.  On the other hand, for other samples where the
photometric errors are large and no spectroscopic redshifts are
available, some model dependence is almost unavoidable.

From a Bayesian perspective, these model dependencies arise as a
dependence on an assumed prior, a problem one has whenever the maximum
likelihood distribution isn't sufficiently narrow.  For a bright HDF
sample where both the flux and redshift uncertainties are small, the
corresponding maximum likelihood distributions are extremely narrow
and so there is minimal dependence on the assumed prior.  For these
cases, there are clear advantages to the present approach where the
base model is comprised of a sample of template galaxies with specific
redshifts (hereafter called the cloning approach) over other
approaches which attempt to model galaxies in a two or three
dimensional space, since we have minimal loss of information.  On the
other hand, in cases where the flux and redshift uncertainties become
large, clearly a lower-dimensional model-dependent Bayesian analysis
becomes more attractive.

Nevertheless, even in the presence of uncertainties there are ways of
making some low-order corrections to the results obtained from the
cloning approach if the data are of low S/N.  We discuss the
corrections in turn.  The first difficulty relates to the redshift
uncertainty of each galaxy template.  Redshift uncertainties result in
a certain smoothing of the intrinsic distribution of sizes and
luminosities, effectively moving the knee of the luminosity function
to higher luminosities.  This becomes problematic when there are
gradients in the density with which particular objects fill redshift
space, as there inevitably are, objects naturally moving from regions
of high redshift space density to regions of low redshift space
density.  One could attempt to correct for this effect by binning the
objects in some way and by making some assumption about their
intrinsic distribution, i.e., the prior, for this class of object, but
in doing so, one would introduce model dependencies.  An alternate
approach for treating this bias would involve \textit{not} correcting
the bias in the original sample itself, but instead introducing a
similar bias in every sample against which one compared the original
sample.  We have employed neither in our paper because our own
simulations have shown that the size of the effect ($<5$\%) is much
smaller than even the Poissonian variation in our small sample.

While it is of course true that uncertainties in the redshift
estimators can bias the predictions of the derived model, a more
worrisome issue is the possibility that systematic uncertainties in
the redshifts derived would not only bias these same predictions, but
bias them severely.  Since the reliability of photometric redshifts
here depends on the extent to which our spectral templates accurately
represent the shape and curvature of the true SEDs, it would be rather
easy to make systematic errors across our high redshift samples.
Fortunately, a large number of our bright galaxies have spectroscopic
redshifts and so the reliability of our photometric redshifts can be
checked, at least at brighter magnitudes (see Figure 4).

There are also uncertainties in the pixel fluxes, and these
uncertainties will affect both our photometry and other parametric
determinations, and such errors will have an effect on whether a
galaxy is selected as part of our sample or not.  This isn't a problem
if galaxies uniformly populate the parameter space over the selection
window since objects will as likely be scattered into a region as out
of it.  Unfortunately, in the more common case where there are
gradients in the way objects fill multi-dimensional parameter space
objects will naturally be scattered from the more dense regions to the
less dense regions (the Malmquist bias being a well-known example of
this).  As per uncertainties in redshift, corrections require binning
the objects in some manner and making some assumption about their
intrinsic distribution or prior.

Not only will the uncertainties in the pixel flux have an effect on
whether an object is selected or not, they will have an effect on how
an object is resimulated at both lower and higher redshift.  For
example, if errors in the pixel fluxes result in an object's seeming
redder or bluer than it really is, then the resimulated object will
consistently appear redder or bluer than it really is, biasing the
determined selection function.  Fortunately, for the present
situation, this has only a $\sim3-5\%$ effect on the determined volume
density, basically because the bulk of the $U$ dropout objects do not
lie very close to the $B-I$ edge of the selection window and therefore
the scatter toward intrinsically bluer/redder colours does not
appreciably change the volume density of sample objects by much on
average.\footnote{See de Jong \& Lacey (2000) for a proposed solution
to this.  We will be attempting to incorporate this procedure into
future applications of this methodology.}

Despite our emphasis on a model-independent cloning approach, a
lower-dimensional parametric approach (discussed above) works well in
many cases, as the analysis that Steidel et al.\ (1999) and Deltorn et
al.\ (2001) make of the $U$ dropouts effectively illustrates.  In both
studies, conclusions about the $U$-dropout population and its relation
to higher redshift populations are made based upon a two-variable
parameterization: the absolute magnitude at 1700 $\AA$ and the
spectral type.  Compared with our work, these analyses have their pros
and cons.  They are better in the sense that allow a more
straightforward treatment of errors in redshift and photometry as
detailed in this section and enable one to push fainter by considering
probable but not certain detections--i.e., Pozzetti et al.\ (1998)'s
pushing fainter than us in their determination of the $U$ and $B$
luminosity functions through their use of Kaplan-Meier (Lavalley,
Isobe, \& Feigelson 1992) estimators.  They also allow a more proper
treatment of objects which are rare enough not to be found in the
input sample used for constructing the empirical models.\footnote{For
this reason, Bouwens, Broadhurst, \& Silk (1998b) included a
low-luminosity galaxy model along with their empirical clone model to
put everything in context.}  They are worse in the sense that they
tend to assume that the joint multivariate distribution of surface
brightness, shape, blochiness, or pixel-by-pixel color variation is
independent of luminosity or spectral type.

\subsection{Object Overlap}

Another challenge regards the overlap or deblending of different
galaxies.  Not only is this a problem when one selects the original
sample, but it is also a problem when one resimulates these objects in
different environments where they may or may not closely overlap with
background/foreground objects.  For the former problem, one approach
is simply to consider these objects as more complicated isolated
galaxies.  Of course, if the contaminating object is at a very
different redshift from the foreground galaxy, attempts at replicating
this object to other redshifts can be a problem, the problem being
more significant the more equal the fluxes of the blended objects, the
more dissimilar their redshifts.  This is just another example of the
problems one faces when attempting to derive an empirical model from
real objects on real images with all the associated uncertainties in
redshifts, pixel fluxes, and possible blending problems.  As for the
latter problem, it is quite naturally treated by adding the simulated
objects to a frame full of foreground and background objects, and
treating the blended objects from the simulations just as one does in
the real samples.  Fortunately, both problems appear to have a very
small effect, affecting only $\sim2$\% of the objects (a rough
estimate based upon the overlapping objects found in the HDFs).  Note
that for speed and simplicity, we have measured the properties of
sample galaxies in isolation.  Not only would a full treatment of
overlap require a lot more computational resources, but it would be
impossible to do exactly right due to difficulties involved in
separating foreground objects from background objects on which they
are superimposed.

\subsection{Inadequate S/N}

Determining the selection window for almost any object requires one to
resimulate the galaxy at both lower and higher redshifts.  When
projected to lower redshifts, the S/N of the template is not generally
good enough for an accurate comparison with the real high S/N
lower-redshift data.  In these cases, we have chosen simply to allow
the noise to increase, resulting in a larger scatter in the recovered
magnitudes.  Obviously, one should pay attention to these issues when
interpreting the data or designing the original selection criteria,
but they do not introduce any significant systematics here.  Another
difficulty occurs in trying to simulate objects at very low redshift
where the angular extent of objects becomes very large relative to the
pixel scale.  This significantly increases the simulation time.  To
speed things up, for cases where the pixel sizes are smaller than the
projected size of template pixels, the pixel sizes and zero-points of
the simulated images are scaled up to match that of the template
pixels.

\end{document}